\documentclass[structabstract]{aa}
\usepackage[modulo,switch]{lineno}
\usepackage{graphicx}
\usepackage{natbib}
\usepackage{longtable}
\usepackage{longtable,lscape}
\usepackage{txfonts}
\begin{document}
\newcommand{\meandnu} {\langle\Delta\nu\rangle}
   \title{Solar-like oscillations in red giants observed with \textit{Kepler}: comparison of global oscillation parameters from different methods}

   \author{S. Hekker\inst{1} \and Y. Elsworth\inst{1} \and J. De Ridder\inst{2} \and B. Mosser\inst{3} \and R.A. Garc\'{i}a\inst{4}  \and T. Kallinger\inst{5}\fnmsep\inst{6} \and S. Mathur\inst{7} \and D. Huber\inst{8} \and D.L. Buzasi\inst{9}  \and H.L. Preston\inst{9}\fnmsep\inst{10} \and S.J. Hale\inst{1} \and J. Ballot\inst{11} \and W.J. Chaplin\inst{1} \and C. R\'egulo\inst{12}\fnmsep\inst{13} \and T.R. Bedding\inst{8} \and D. Stello\inst{8} \and W.J. Borucki\inst{14} \and D.G. Koch\inst{14} \and J. Jenkins\inst{15} \and C. Allen\inst{16} \and R.L. Gilliland\inst{17} \and H. Kjeldsen\inst{18} \and J. Christensen-Dalsgaard\inst{18}}

\offprints{S. Hekker, \\
                    email: saskia@bison.ph.bham.ac.uk}
   \institute{University of Birmingham, School of Physics and Astronomy, Edgbaston, Birmingham B15 2TT, United Kingdom
   \and  Instituut voor Sterrenkunde, K.U. Leuven, Celestijnenlaan 200D, 3001 Leuven, Belgium
   \and LESIA, UMR8109, Universit\'e Pierre et Marie Curie, Universit\'e Denis Diderot, Observatoire de Paris, 92195 Meudon Cedex, France
   \and Laboratoire AIM, CEA/DSM-CNRS, Universit\'{e} Paris 7 Diderot, IRFU/SAp, Centre de Saclay, 91191, GIf-sur-Yvette, France
   \and Department of Physics and Astronomy, University of British Colombia, 6224 Agricultural Road, Vancouver, BC V6T 1Z1, Canada
   \and Institute for Astronomy, University of Vienna, T\"urkenschanzstrasse 17, A-1180 Vienna
   \and High Altitude Observatory, NCAR, P.O. Box 3000, Boulder, CO 80307, USA
   \and Sydney Institute for Astronomy (SIfA), School of Physics, University of Sydney, NSW 2006, Australia
   \and Eureka Scientific, 2452 Delmer Street Suite 100, Oakland, CA 94602-3017, USA
   \and Department of Mathematical Sciences, University of South Africa, Box 392, UNISA 0003, South Afrika
   \and Laboratoire d'Astrophysique de Toulouse-Tarbes, Universit\'e de Toulouse, CNRS, F-31400, Toulouse, France
   \and Universidad de La Laguna, Dpto de Astrof\'isica, 38206, Tenerife, Spain
   \and Instituto de Astrof\'\i sica de Canarias, 38205, La Laguna, Tenerife, Spain
   \and NASA Ames Research Center, MS 244-30, Moffet Field, CA 94035, USA
   \and SETI Institute, NASA Ames Research Center, MS 244-30, Moffet Field, CA 94035, USA
   \and Orbital Sciences Corp., NASA Ames Research Center, Moffet Field, CA 94035, USA
   \and Space Telescope Science Institute, 3700 San Martin Drive, Baltimore, MD 21218, USA
   \and Department of Physics and Astronomy, Building 1520, Aarhus University, 8000 Aarhus C, Denmark\\
         }

   \date{Received ; accepted}


  \abstract
   {The large number of stars for which uninterrupted high-precision photometric timeseries data are being collected with \textit{Kepler} and CoRoT initiated the development of automated methods to analyse the stochastically excited oscillations in main-sequence, subgiant and red-giant stars.}
   {We investigate the differences in results for global oscillation parameters of G and K red-giant stars due to different methods and definitions. We also investigate uncertainties originating from the stochastic nature of the oscillations.}
   {For this investigation we use \textit{Kepler} data obtained during the first four months of operation. These data have been analysed by different groups using already published methods and the results are compared. We also performed simulations to investigate the uncertainty on the resulting parameters due to different realizations of the stochastic signal.}
   {We obtain results for the frequency of maximum oscillation power ($\nu_{\rm max}$) and the mean large separation ($\meandnu$) from different methods for over one thousand red-giant stars. The results for these parameters agree within a few percent and seem therefore robust to the different analysis methods and definitions used here. The uncertainties for $\nu_{\rm max}$ and $\meandnu$ due to differences in realization noise are not negligible and should be taken into account when using these results for stellar modelling.
   }
   {
   }

   \titlerunning{Comparison of global oscillation parameters from different methods}
   \authorrunning{Hekker et al.}
   \maketitle
%

\section{Introduction}
Oscillations in stars provide information on their internal structures. In the Sun, oscillations are stochastically excited in the outer turbulent layer and these so-called solar-like oscillations have provided a detailed picture of its internal structure including the size of the core, the location of the tachocline and differential rotation \citep[see for a recent overview][]{chaplin2008}. With asteroseismology we observe stellar oscillations to reveal similar information on the internal structures of stars other than the Sun.

Stars with turbulent outer layers are expected to exhibit solar-like oscillations and these are indeed observed in solar-type stars, subgiants and G-K red-giant stars. For recent overviews of these observations and interpretations see e.g., \citet{bedding2008,hekker2010c}. 

Currently, long time series of uninterrupted high-precision photometric data of oscillating stars are being measured by the space missions CoRoT \citep[Convection Rotation and planetary Transits,][]{baglin2006} and \textit{Kepler} \citep[][see also the MAST website http://stdatu.stsci.edu/kepler/ for Kepler data products available for general public]{borucki2009}. These missions have increased spectacularly the number of G and K red-giant stars in which solar-like oscillations, both radial and non-radial,  are clearly detected. First results on red-giant seismology from both missions were presented by e.g. \citet{deridder2009,hekker2009,miglio2009,carrier2010,kallinger2010,mosser2010,bedding2010a,hekker2010b,stello2010}. 

In general, the time series are processed to give information on their frequency content using Fourier methods that take account of the non-uniform sampling in the time domain. When analysing these Fourier spectra, one can distinguish between an approach of detailed modelling of individual oscillation modes using Lorentz fitting \citep[e.g.,][]{appourchaux2008,benomar2009,deheuvels2010}, and a more global analysis of the power spectrum which yields average seismic quantities such as the frequency of maximum power ($\nu_{\rm max}$), the mean large frequency spacing ($\meandnu$), and a parametric fit of the granulation background. From these global seismic observables astrophysically interesting information can be extracted, such as constraints on our galaxy's starburst history \citep{miglio2009} and a precise determination of red
giant radii and masses \citep[e.g.,][]{kallinger2010}. 

However, as will be shown in Sect. 2, different authors use different definitions and different approximations to derive $\nu_{\rm max}$ and $\meandnu$. None of these approaches can be claimed as `the' only correct one, yet there is a need to convey results to theoreticians for modelling. In this paper we quantitatively investigate to what extent the results from different approaches differ. We also study how robust are the different definitions against variations in the power spectrum due to the stochastic realization of the oscillation modes. Understanding both these sources of uncertainty in the observations is of importance for the physical interpretation of the results and seismic modelling. Finally, we discuss what can be gained by combining different methods. 

The results for the global oscillation parameters of stars observed with \textit{Kepler} as presented here are used for a more detailed study by \citet{huber2010} and to investigate the asteroseismic masses and radii of the stars by \citet{kallinger2010b}.

\section{Oscillation parameters}
The particular oscillation parameters discussed in this study are the frequency of maximum oscillation power ($\nu_{\rm max}$) and the large frequency separation ($\Delta \nu$). In general, $\Delta \nu$ is the separation between oscillation modes with the same degree $\ell$ and consecutive radial orders $n$ and is sensitive to the sound travel time across the star. For high-order low-degree modes we expect the eigenfrequencies ($\nu_{n,\ell}$) to follow approximately the asymptotic relation by \citet{tassoul1980}, which can be expressed as:
\begin{equation}
\nu_{n,\ell} \simeq \Delta \nu \left(n+{\ell \over 2} + \varepsilon\right)-\ell(\ell+1)D_0,
\label{asymptot}
\end{equation}
in which $\varepsilon$ is sensitive to the surface layers and $D_0$ to layers deeper inside the star. Based on previous results, we take $D_0$ to be small compared to $\Delta \nu$. The large separation is approximately constant. However, it is well known from observations of the Sun and other stars that $\Delta \nu$ does depend on both frequency and angular degree. 

We first describe the general methods by which we obtain $\nu_{\rm max}$ and $\meandnu$. This will then be followed by a more detailed description of each method used in the present study, with the aim of investigating the similarities and differences in the resulting values obtained from different determinations.

The presence of modes due to resolved, stochastically-excited solar-like oscillations leads to a power excess  which has a shape that can often be approximated by a Gaussian distribution. This characteristic feature is used to estimate $\nu_{\rm max}$ as the centroid of a Gaussian fitted to the (smoothed) power excess. A complexity is that the oscillations in red giants occur in a frequency regime also containing signal due to stellar activity, granulation and possibly instrumental effects. This `background' power has a slope that can influence the determination of $\nu_{\rm max}$, and must be taken into account. The background can be modeled by a sum of power laws and white noise \citep[e.g.][]{harvey1985,hekker2009,kallinger2010}. In some cases a simplification is valid and a linear approximation to the background over the frequency range of the oscillations is used.

The regularity in the large frequency spacing between the modes of solar-like oscillations  provides a clear signature in the power spectrum of the power spectrum (PS$\otimes$PS), which is equivalent to the autocorrelation of the timeseries. The highest features in the PS$\otimes$PS occur at submultiples of $\Delta \nu$, i.e., $\Delta \nu/2$, $\Delta \nu/4$. Similar features from submultiples of $\Delta \nu$ are present in an autocorrelation of the power spectrum. Many algorithms use a rough version of the relation between $\nu_{\rm max}$ and $\Delta \nu$ as described by \citet{hekker2009,stello2009,mosser2010} to identify which feature is due to which submultiple:
\begin{equation}
\Delta \nu \sim \nu_{\rm max}^{0.73 \to 0.8},
\label{dnunumax}
\end{equation}
in which $0.73 \to 0.8$ indicates the range of values used for the exponent.
When the individual modes in the power spectrum are fitted directly, in a procedure known as peak bagging, the large separation as a function of frequency can be obtained directly. Peakbagging is only used in one approach, which is described in more detail by \citet{kallinger2010b}.

Having described the general concepts for determining $\nu_{\rm max}$ and $\meandnu$, we now describe the methods used by the different teams that analysed the data. The methods are also summarised in Table~\ref{methods}. We refer to the references for more detailed descriptions of the methods developed by the individual teams.

\begin{table*}
\begin{minipage}{\linewidth}
\caption{Summary of the different methods used to determine $\nu_{\rm max}$ and $\meandnu$. We also mention the way the smoothing is applied to the power spectrum in the determination of $\nu_{\rm max}$, the frequency interval over which $\meandnu$ is computed, how the background is determined and whether Eq.~\ref{dnunumax} is used. For the background we indicate the general formulae with $w$ indicating white noise, $\nu$ frequency, and $a, B, C, \alpha$ and $\beta$ free parameters.}
\label{methods}
\centering
\begin{tabular}{l|cccccc}
\hline\hline
ID & COR & CAN & A2Z & SYD & DLB & OCT \\
\hline
$\nu_{\rm max}$  & centroid Gaussian& centroid Gaussian & centroid Gaussian & peak smoothed PS & centroid Gaussian & centroid Gaussian (I)\\
 & & & & & & first moment (II)\\
smoothing & 3$\Delta\nu$ Gauss & none & 4$\Delta\nu$ box & 2$\Delta\nu$ Gauss & 10 $\mu$Hz Gauss & 2$\Delta\nu$ box\\
$\Delta \nu$ & autocor timeseries & fit to PS & PS$\otimes$PS & autocor PS & autocor PS & PS$\otimes$PS \\
freq interval & FWHM osc excess & 3 radial orders & $\nu_{\rm max} \pm \nu_{\rm max}/3$ & $\nu_{\rm max} \pm$ 5-7$\Delta\nu$ & $\nu_{\rm max} \pm \nu_{\rm max}/4$ & osc excess\\
background & $w + a\nu^{-\beta}$ & $w + \sum_{i=1}^{3}\frac{B_i}{1+C_i\nu^4}$ & $w + \frac{B}{1+C\nu^{\alpha}} + a\nu^{-\beta}$ & $w + \sum_i\frac{B_i}{1+(C_i\nu)^2 +(C_i\nu)^4}$ & $w + \frac{B}{1+C\nu^2}$  & linear approx\\
Eq.~\ref{dnunumax} & yes & no & yes & yes & yes & yes \\
\hline
\end{tabular}
\end{minipage}
\end{table*}

\begin{itemize}
\item In the autocorrelation method described by \citet[][hereafter COR]{mosser2009} $\nu_{\rm max}$ is obtained as the centroid of a Gaussian fit to the smoothed power spectrum. For the smoothing, a Gaussian with full width half maximum (FWHM) of 3$\meandnu$ is used. $\meandnu$ is determined from the first peak in the autocorrelation of the timeseries apodised with a Hanning filter, where the FWHM of the excess envelope is used to compute the mean value.
\item In the automated Bayesian Markov-Chain Monte Carlo algorithm \citep[][hereafter CAN]{kallinger2010,gruberbauer2009}, $\nu_{\rm max}$ is defined as the centroid of a Gaussian fitted to the unsmoothed power spectrum. $\meandnu$ is obtained from peakbagging, i.e. from fitting a sequence of Lorentzian profiles spanning three radial orders to the background-corrected power spectrum, parameterised by the large and small frequency separations.
\item In the method described by \citet[][hereafter A2Z]{mathur2010} and adapted for the analysis of red giants, $\nu_{\rm max}$ is defined  from a Gaussian fit to the smoothed power spectrum. For the smoothing a boxcar of width 4$\meandnu$ is used. $\meandnu$ is determined from the highest feature in the PS$\otimes$PS, which is computed over a range $\nu_{\rm max} \pm \nu_{\rm max}/3$. The results for $\meandnu$ are cross-checked with results from the autocorrelation method \citep{mosser2009}.
\item \citet[][hereafter SYD]{huber2009} compute $\nu_{\rm max}$ as the peak of the smoothed power spectrum. The smoothed power spectrum is computed using a Gaussian with a FWHM of 2$\meandnu$.  $\meandnu$ is computed from the autocorrelation of the power spectrum in the frequency interval $\nu_{\rm max}$ $\pm$ 5-7$\meandnu_{\rm exp}$, where $\meandnu_{\rm exp}$ is calculated using Eq.~\ref{dnunumax}.
\item In the next method (Buzasi, Preston; hereafter DLB), $\nu_{\rm max}$ is determined from a Gaussian fit to the smoothed power spectrum. The power spectrum is smoothed using a Gaussian filter with a width equal to 10~$\mu$Hz. $\meandnu$ is computed from the autocorrelation of the power spectrum in the range $\nu_{\rm max} \pm \nu_{\rm max}/4$.
\item  In the method described by \citet{hekker2010a} and adapted for red giants, $\nu_{\rm max}$ is determined in two ways. In the first method  (OCT I), $\nu_{\rm max}$ is defined as the centroid of a Gaussian fit to the smoothed power spectrum. In the second method (OCT II), $\nu_{\rm max}$ is computed as the first moment of the area under the smoothed power envelope. The smoothing is obtained using a boxcar with a width of 2$\meandnu$. $\Delta \nu$ is computed from the PS$\otimes$PS of the full frequency range in which oscillation excess has been detected.
\end{itemize}

\begin{figure}
\begin{minipage}{\linewidth}
\centering
\includegraphics[width=\linewidth]{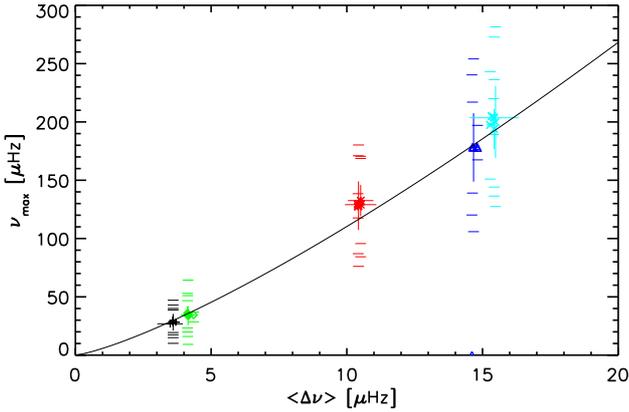}
\end{minipage}
\caption{$\nu_{\rm max}$ versus $\meandnu$ for five stars observed with \textit{Kepler}. For each star all results are plotted with the uncertainties derived by the different methods. The ranges in frequency over which $\meandnu$ values are computed are indicated with the horizontal lines above and below the results. A different colour and symbol are used for  results of each star. The black solid line indicates the approximate relation between $\meandnu$ and $\nu_{\rm max}$.}
\label{dnurange}
\end{figure}

\noindent From this overview it is clear that similar approaches are used by the different teams to determine $\nu_{\rm max}$ and $\meandnu$, although there are differences in the smoothing for $\nu_{\rm max}$ and in the range over which the mean $\Delta \nu$ is computed. On average, neither of these effects are significant in comparison with other sources of uncertainty. We illustrate this for $\meandnu$ with the data shown in Fig.~\ref{dnurange}, which shows the results from several methods for $\nu_{\rm max}$ and $\meandnu$ for five typical stars. The observed values are shown by the symbols with the ranges over which $\meandnu$ has been computed indicated by the small horizontal lines. Formal errors, where available, are also shown. It is obvious from this figure that the impact of the frequency range is small and within the uncertainties. 
This is plausible for the following reasons. Firstly, the trend in $\Delta\nu$ over a typical frequency range is approximately linear with frequency. Note that, when analysed in detail, it can be shown that $\Delta\nu$ may vary rapidly with frequency \citep{mosser2010}. Secondly, most of the time, the region selected is symmetrical with respect to $\nu_{\rm max}$. The consequence of these two factors is that the value returned for $\meandnu$ is roughly independent of the precise range chosen. Furthermore, $\Delta\nu$ changes relatively slowly with frequency and so the impact of a small change in the central position is small, i.e., within the general uncertainties in the determination of $\meandnu$.

Differences in results for $\nu_{\rm max}$ and $\meandnu$ may also arise from the following issues:
\begin{itemize}
\item the preparation of the power spectrum,
\item identification of the frequency range of oscillations,
\item computation of the background signal,
\item definition of a positive detection / artefact,
\item computation of uncertainties.
\end{itemize}
These issues are addressed in Sects. 4 and 5.

\section{Comparison using simulations}
To obtain a measure of the variation in the observed parameters due to the stochastic nature of the oscillations, which we will refer to as realization noise, we first present results of a comparison using different realizations of simulated timeseries. Because we observe stochastically excited oscillations, the observed oscillation characteristics will change with time. Timeseries with an infinite timespan would provide the `real' or `limit' oscillation parameters.
In reality we are dealing with data with a limited timespan and to investigate the scatter in the results due to this, we simulated 100 realizations for five `\textit{Kepler}' stars using the simulator described by \citet{deridder2006b}. The input parameters are listed in Table~\ref{input} and were derived from the observations of the \textit{Kepler} targets with \textit{Kepler} ID 1720425, 2013502, 2696732, 3526061, and 6033938 (see Table~\ref{results} for their characteristics). These stars are selected based on a visual inspection of their power spectra. By taking parameters obtained from observed stars instead of from for instance scaling relations, we aim to resemble the Kepler observations as well as possible. This also means that scaling relations are not readily applicable anymore, because stars with for instance different metallicity and/or mass are likely to be represented. These parameters are known to induce deviations from the scaling relations. As indicated in Table~\ref{input}, two sets of mode life times have been considered for each of the simulated \textit{Kepler} stars. This was done to investigate the influence of the width of the frequency peaks, which increases for modes with decreasing life times, on the resulting values of $\nu_{\rm max}$ and $\meandnu$. 

\begin{table*}
\begin{minipage}{\linewidth}
\caption{Input parameters for synthetic timeseries: frequency of maximum oscillation power ($\nu_{\rm max}$), large separation between modes of consecutive order ($\Delta \nu$), small separation between modes with degree 0 and 2 ($\delta \nu_{02}$), half width at half maximum of a Gaussian fit to the oscillation power excess (width$_{\rm env}$) in $\mu$Hz, the height of this Gaussian fit (height$_{\rm env}$) to the power excess envelope in ppm$^2$$\mu$Hz$^{-1}$ and mode life times ($\tau$) for modes of different degree $\ell$ = (0,1,2,3) in days.}
\label{input}
\centering
\begin{tabular}{cccccccccc}
\hline\hline
Simulation & $\nu_{\rm max}$ & $\Delta \nu$ & $\delta \nu_{02}$ & width$_{\rm env}$ & height$_{\rm env}$ & $\tau_{\ell = 0}$ & $\tau_{\ell = 1}$ & $\tau_{\ell = 2}$ & $\tau_{\ell = 3}$\\
 & $\mu$Hz  & $\mu$Hz & $\mu$Hz & $\mu$Hz & 10$^3$ ppm$^2$ $\mu$Hz$^{-1}$ & days & days & days & days \\
\hline
1 & 20.0 & 2.7 & 0.38 & 5.0 & 40.0 & 50.0 & 15.0 & 30.0 & 30.0\\
2 & 50.0 & 6.5 & 0.80 & 6.5 & 10.0 & 50.0 & 15.0 & 30.0 & 30.0\\
3 & 80.0 & 9.1 & 1.2   & 6.5 & 14.4 & 50.0 & 15.0 & 30.0 & 30.0\\
4 & 120.0 & 11.2 & 1.4 & 5.0 & 19.6 & 50.0 & 15.0 & 30.0 & 30.0\\
5 & 170.0 & 13.9 & 1.7 & 7.0 & 6.4 & 50.0 & 15.0 & 30.0 & 30.0\\
6 & 20.0 & 2.7 & 0.38 & 5.0 & 40.0 & 100.0 & 30.0 & 60.0 & 60.0\\
7 & 50.0 & 6.5 & 0.80 & 6.5 & 10.0 & 100.0 & 30.0 & 60.0 & 60.0\\
8 & 80.0 & 9.1 & 1.2   & 6.5 & 14.4 & 100.0 & 30.0 & 60.0 & 60.0\\
9 & 120.0 & 11.2 & 1.4 & 5.0 & 19.6 & 100.0 & 30.0 & 60.0 & 60.0\\
10 & 170.0 & 13.9 & 1.7 & 7.0 & 6.4 & 100.0 & 30.0 & 60.0 & 60.0\\
\hline
\end{tabular}
\end{minipage}
\end{table*}

\begin{figure*}
\begin{minipage}{5.6 cm}
\centering
\includegraphics[width=5.6cm]{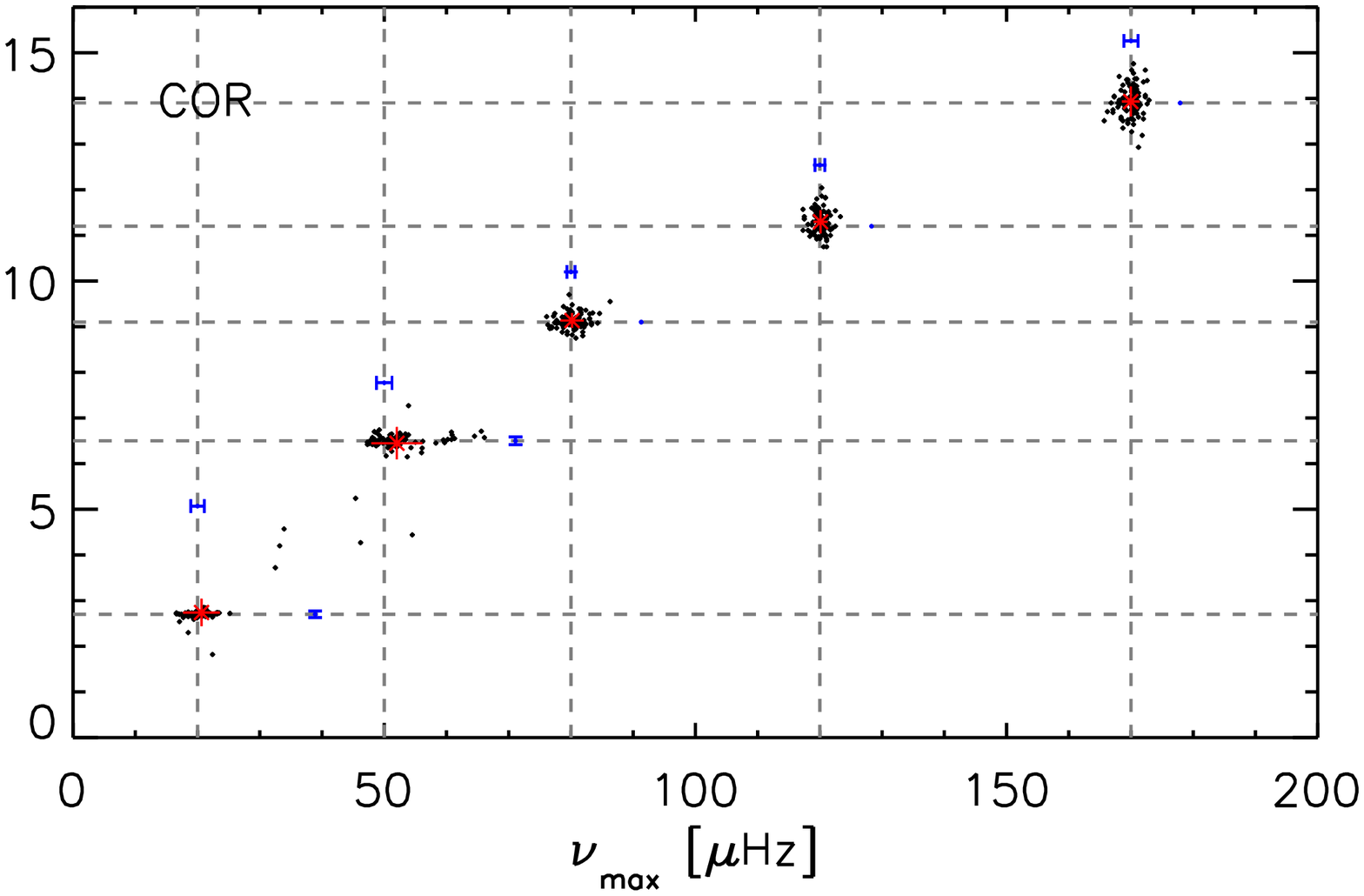}
\end{minipage}
\hfill
\begin{minipage}{5.6 cm}
\centering
\includegraphics[width=5.6cm]{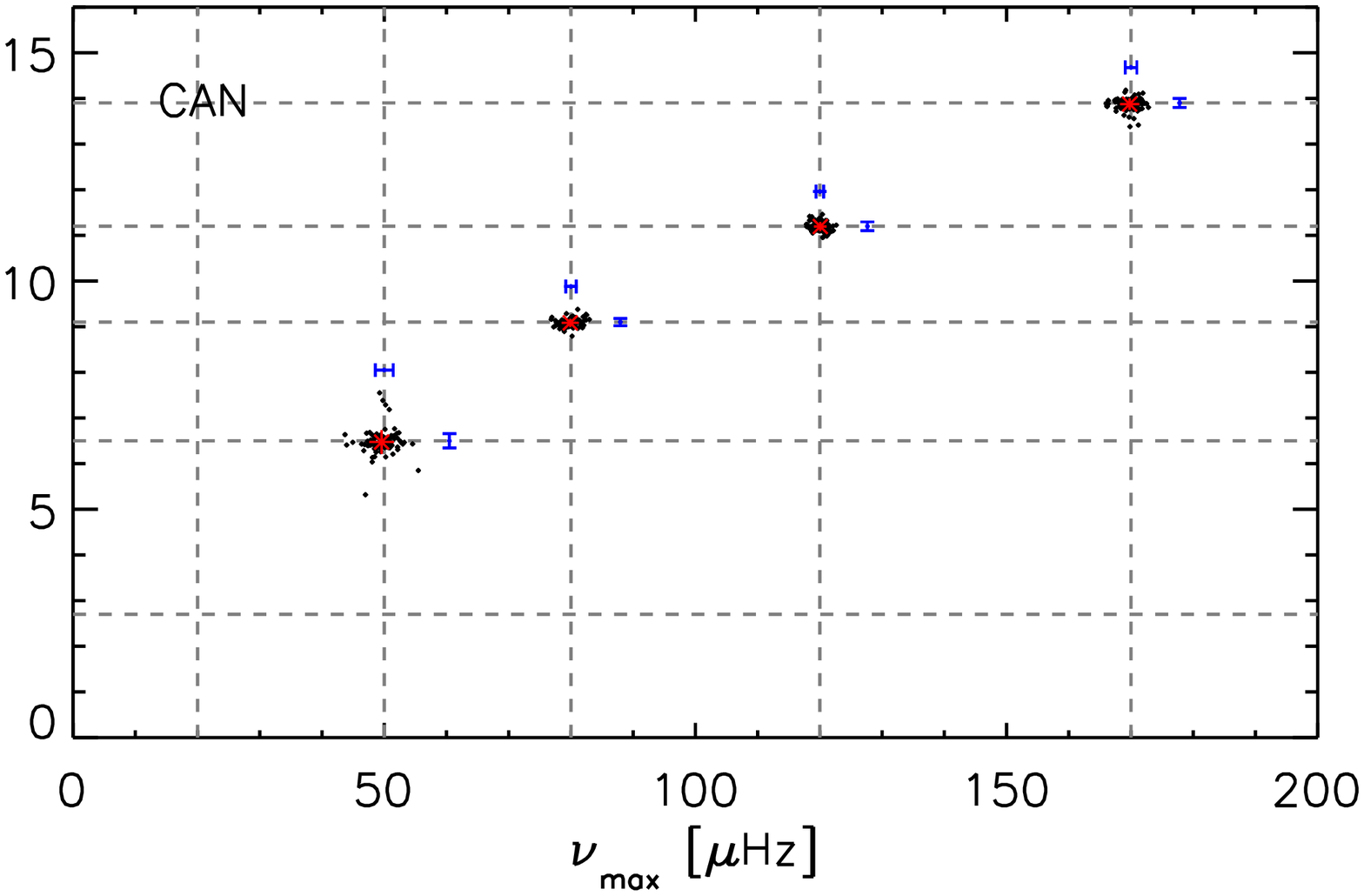}
\end{minipage}
\hfill
\begin{minipage}{5.6cm}
\centering
\includegraphics[width=5.6cm]{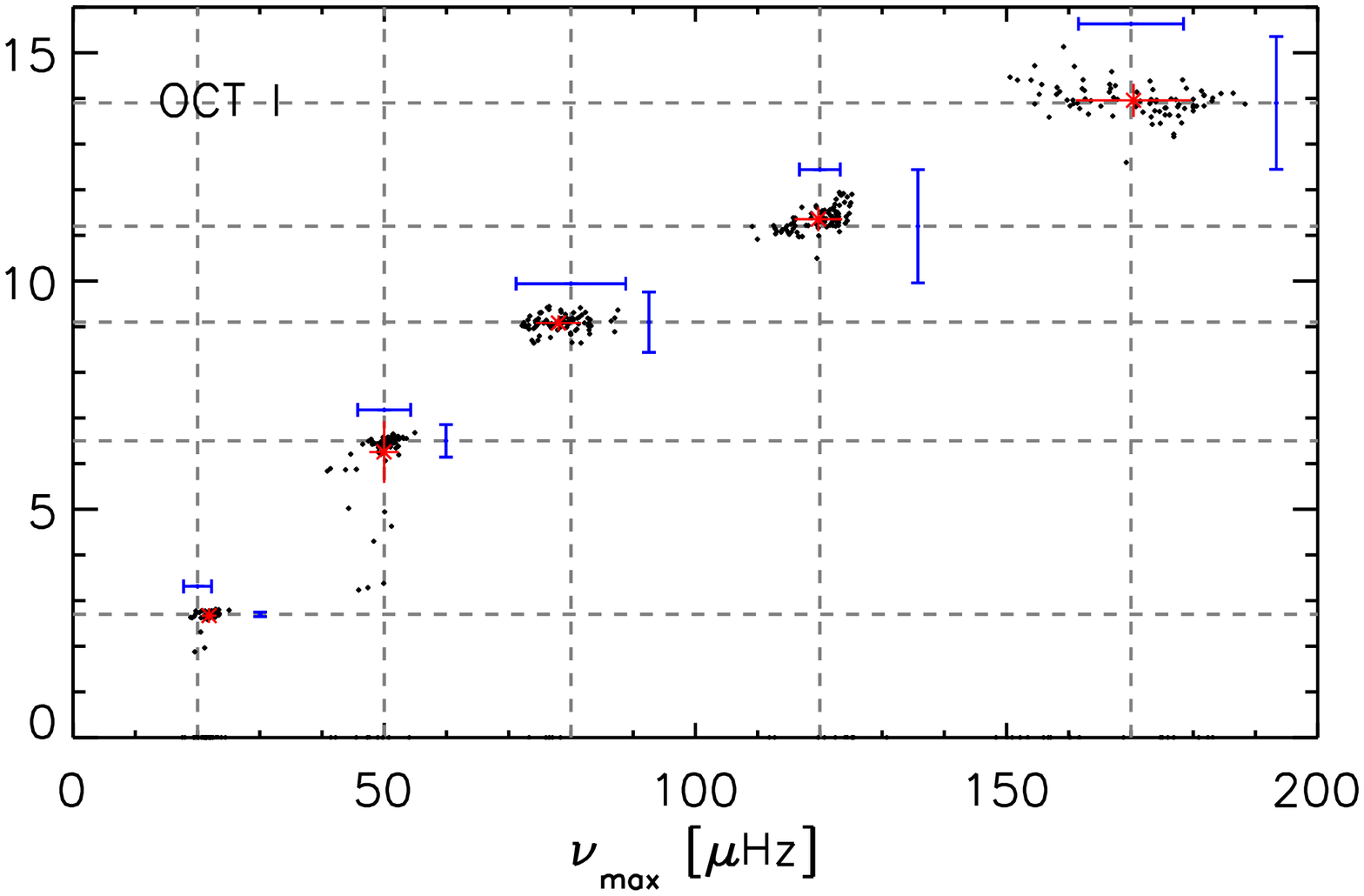}
\end{minipage}
\hfill
\begin{minipage}{5.6cm}
\centering
\includegraphics[width=5.6cm]{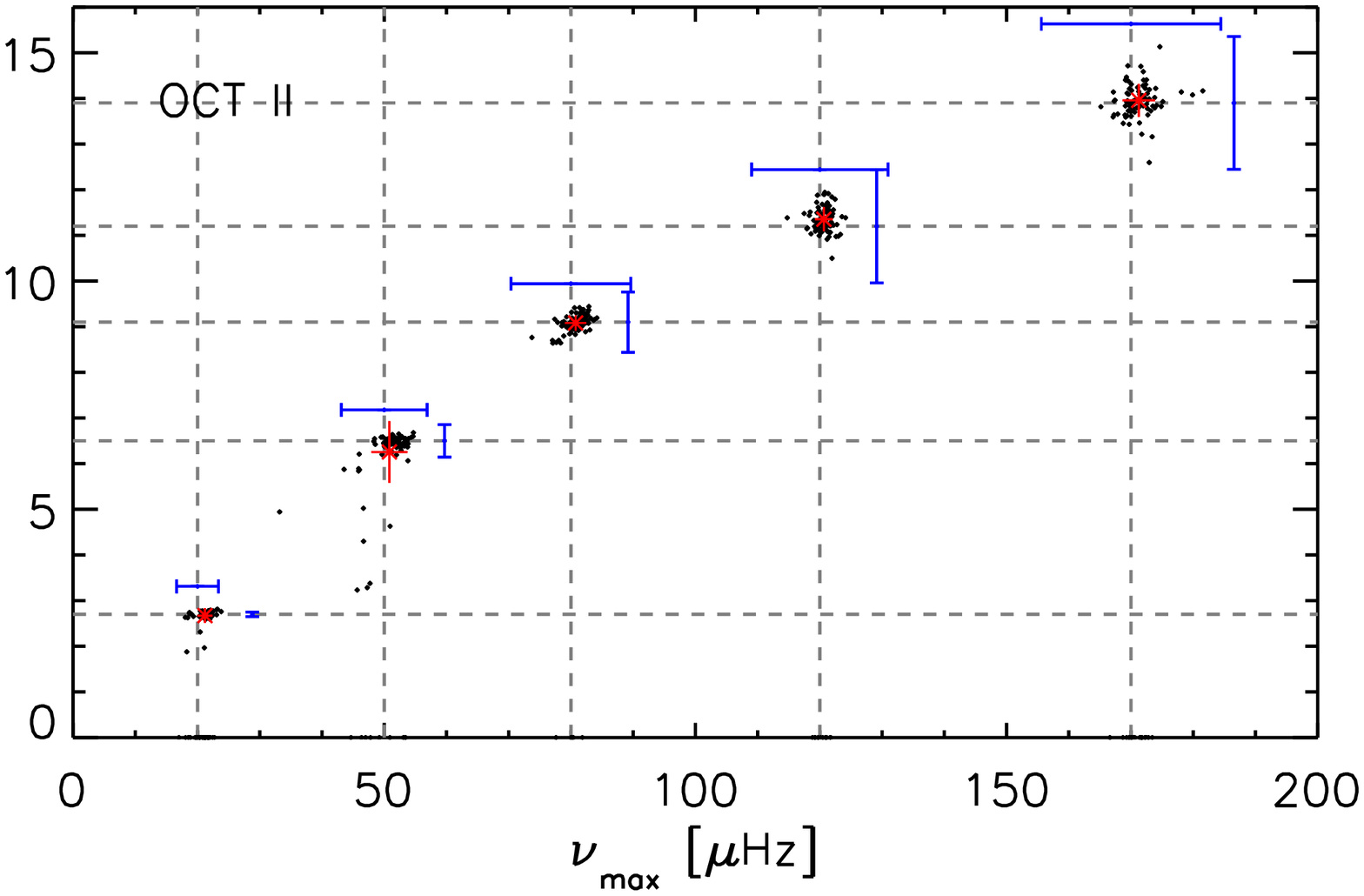}
\end{minipage}
\hfill
\begin{minipage}{5.6cm}
\centering
\includegraphics[width=5.6cm]{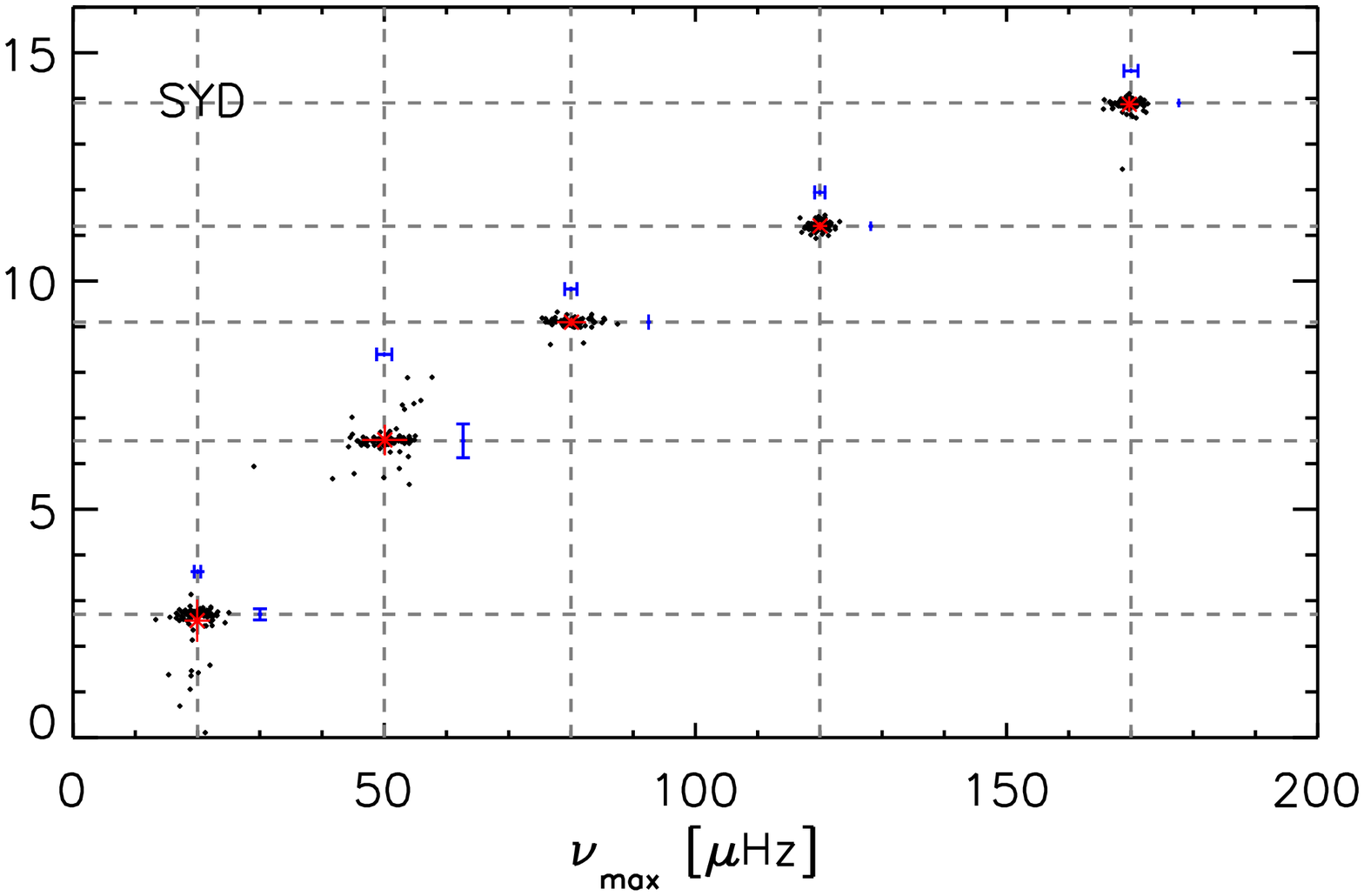}
\end{minipage}
\hfill
\begin{minipage}{5.6cm}
\centering
\includegraphics[width=5.6cm]{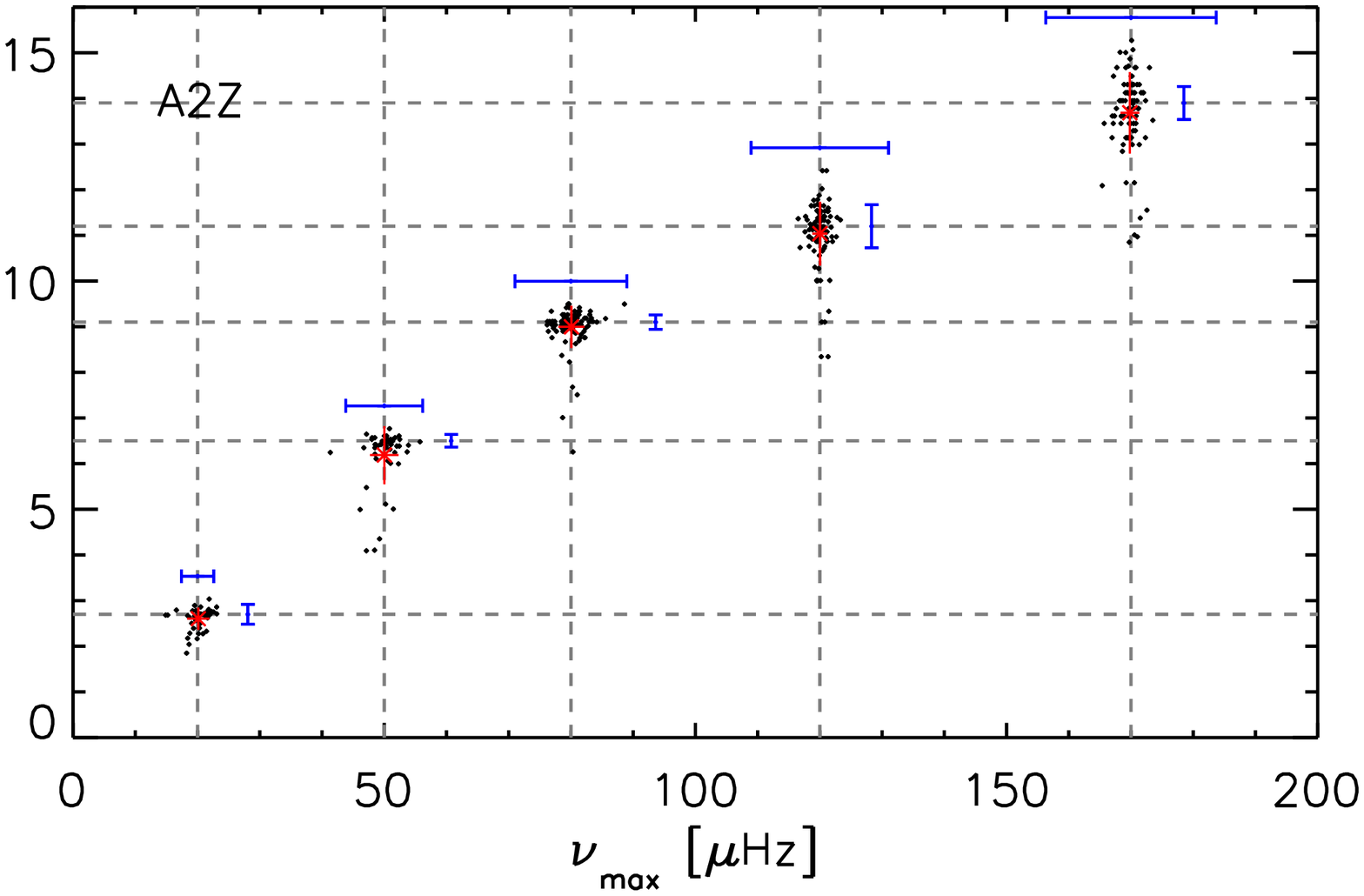}
\end{minipage}
\hfill
\begin{minipage}{5.6cm}
\centering
\includegraphics[width=5.6cm]{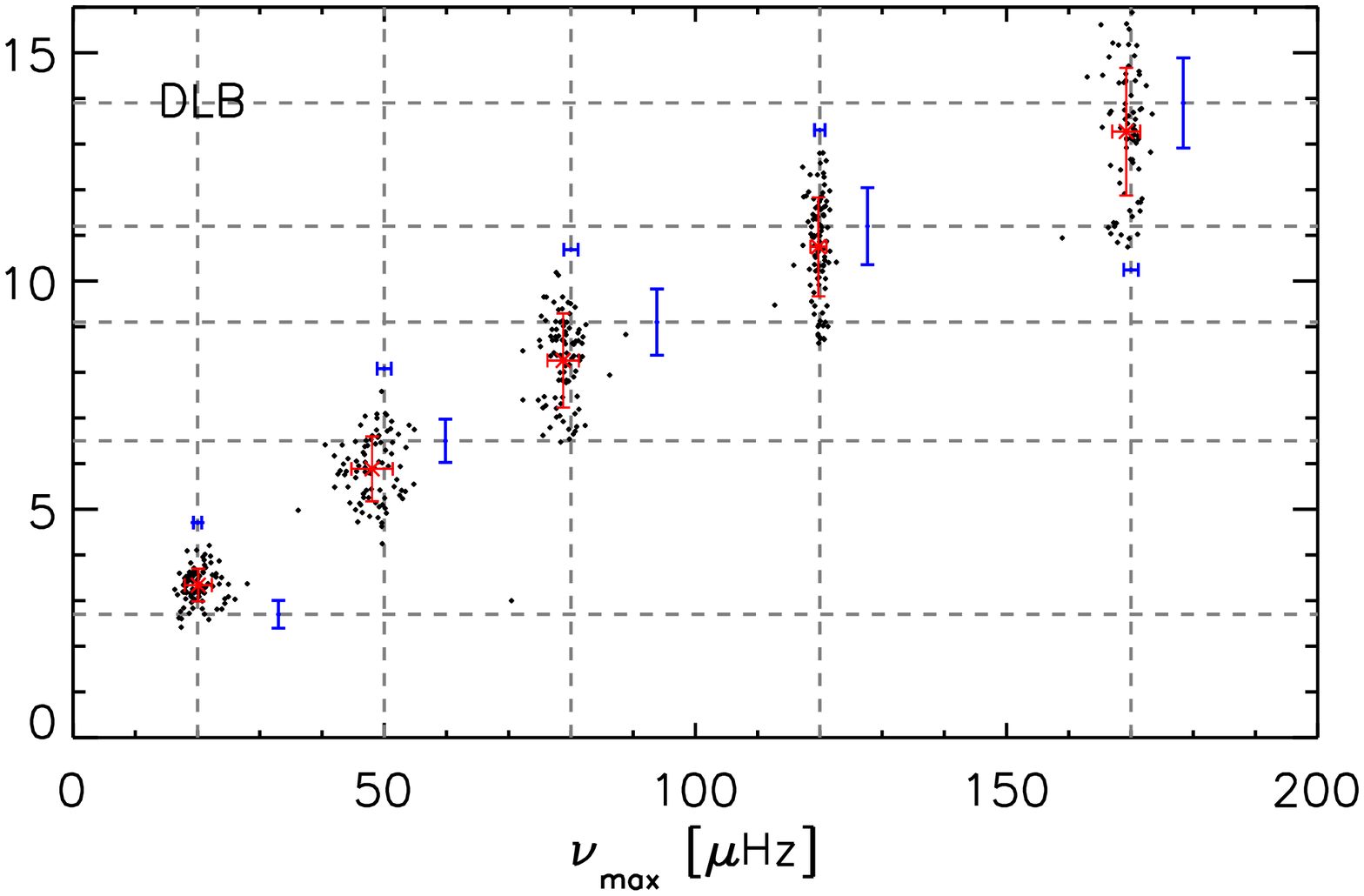}
\end{minipage}
\caption{Results  for  $\nu_{\rm max}$ and $\Delta \nu$ of the analysis of synthetic time series using different methods: COR, CAN, OCT I, OCT II, SYD, A2Z and DLB. The black dots indicate the values obtained by the methods for each individual realization, the red asterisks indicate the mean value and scatter of these results. The typical uncertainties on the computed values for each model are indicated with the blue error bars. The gray dashed lines indicate the input values.} 
\label{ressimnoerror}
\end{figure*}

\begin{figure*}
\begin{minipage}{5.6cm}
\centering
\includegraphics[width=5.6cm]{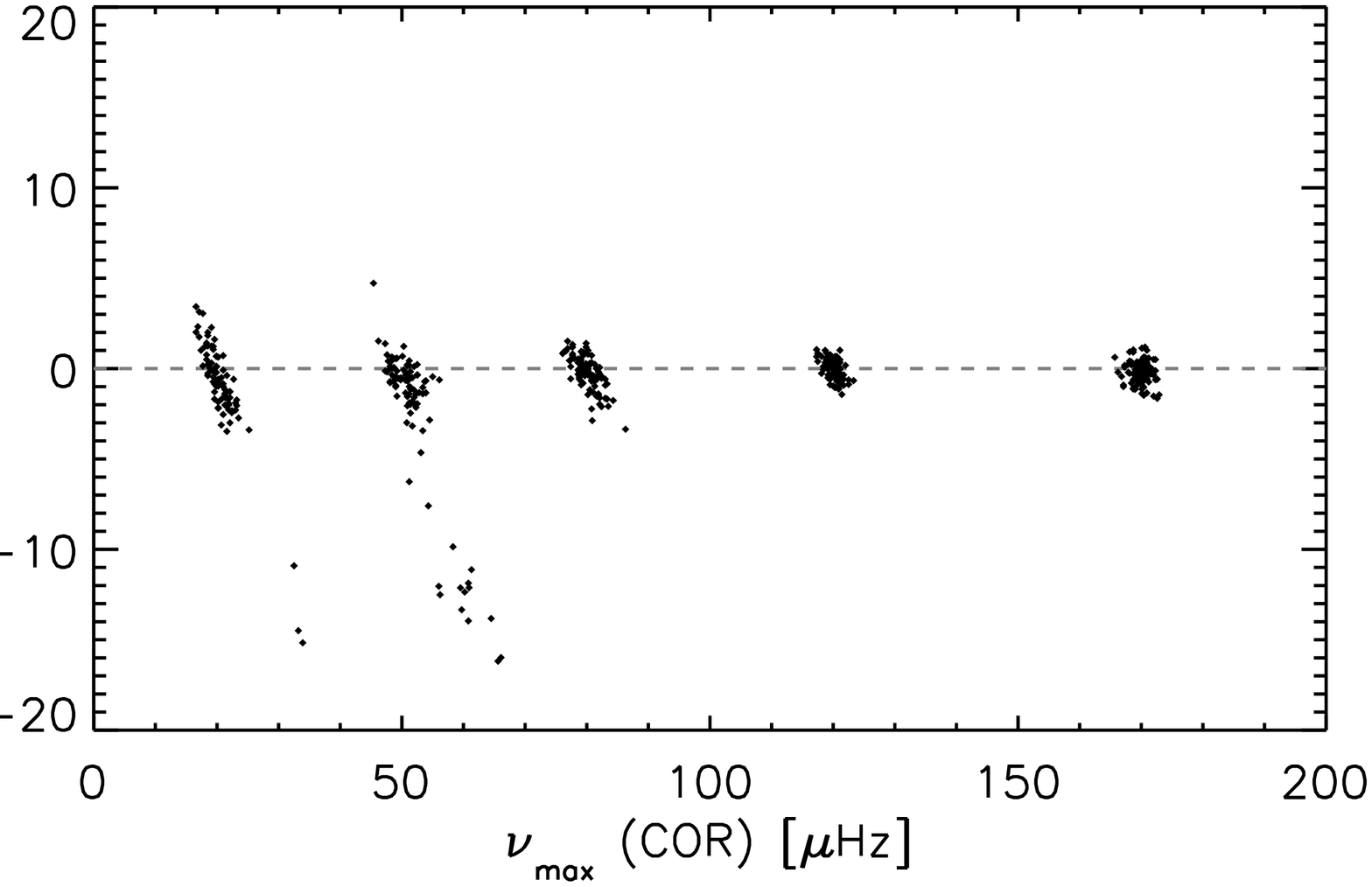}
\end{minipage}
\hfill
\begin{minipage}{5.6cm}
\centering
\includegraphics[width=5.6cm]{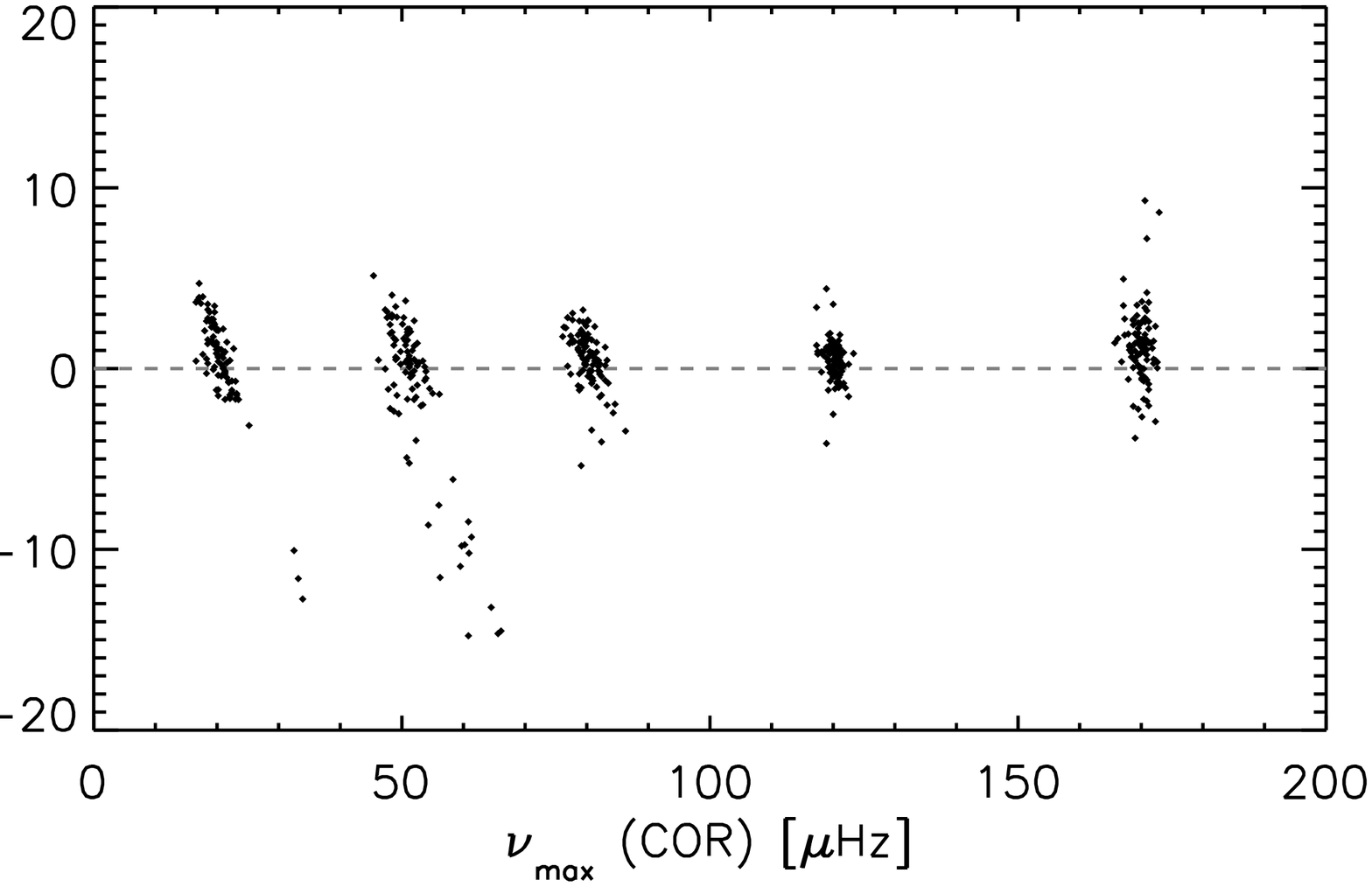}
\end{minipage}
\hfill
\begin{minipage}{5.6cm}
\centering
\includegraphics[width=5.6cm]{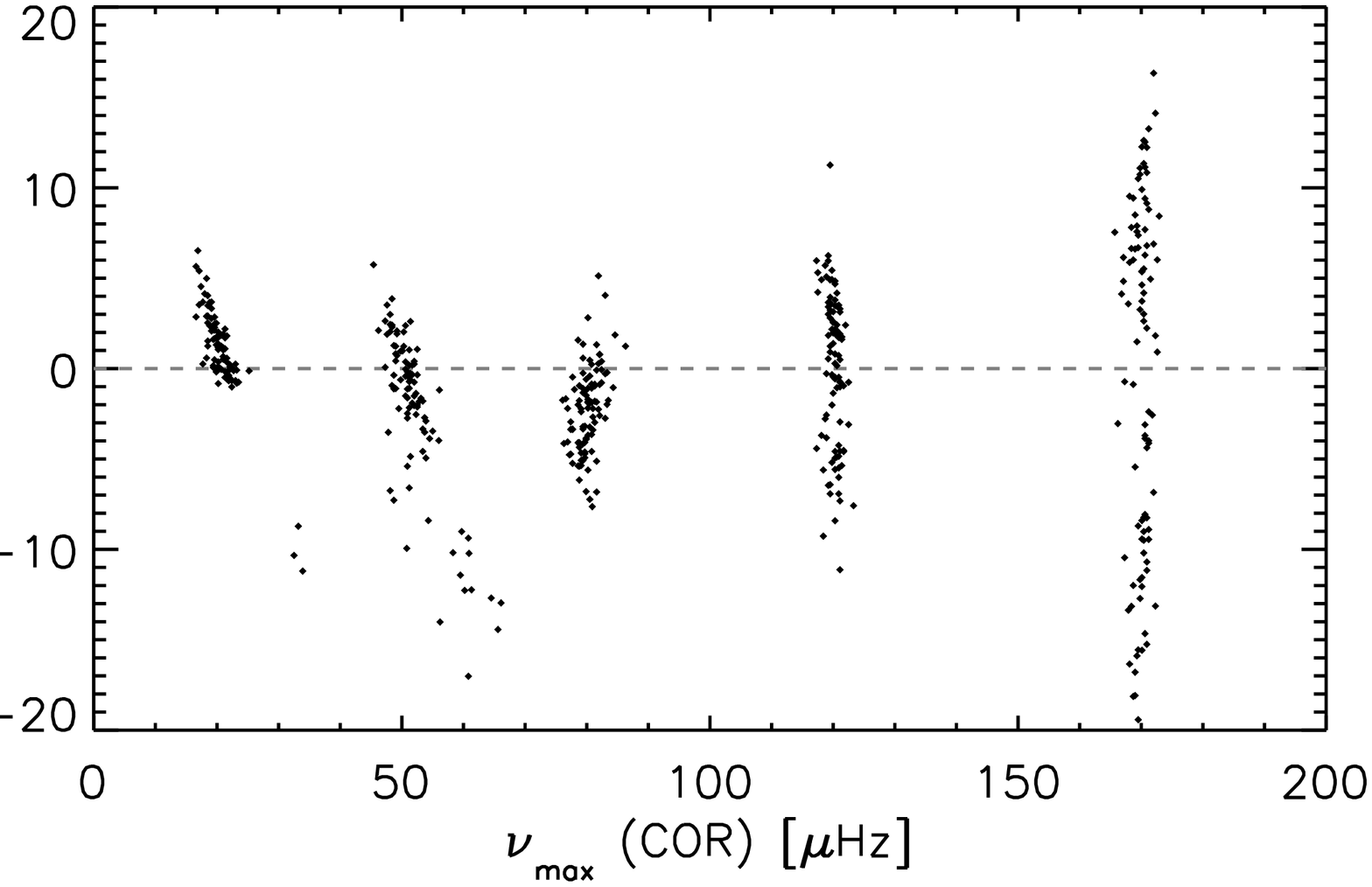}
\end{minipage}
\hfill
\begin{minipage}{5.6cm}
\centering
\includegraphics[width=5.6cm]{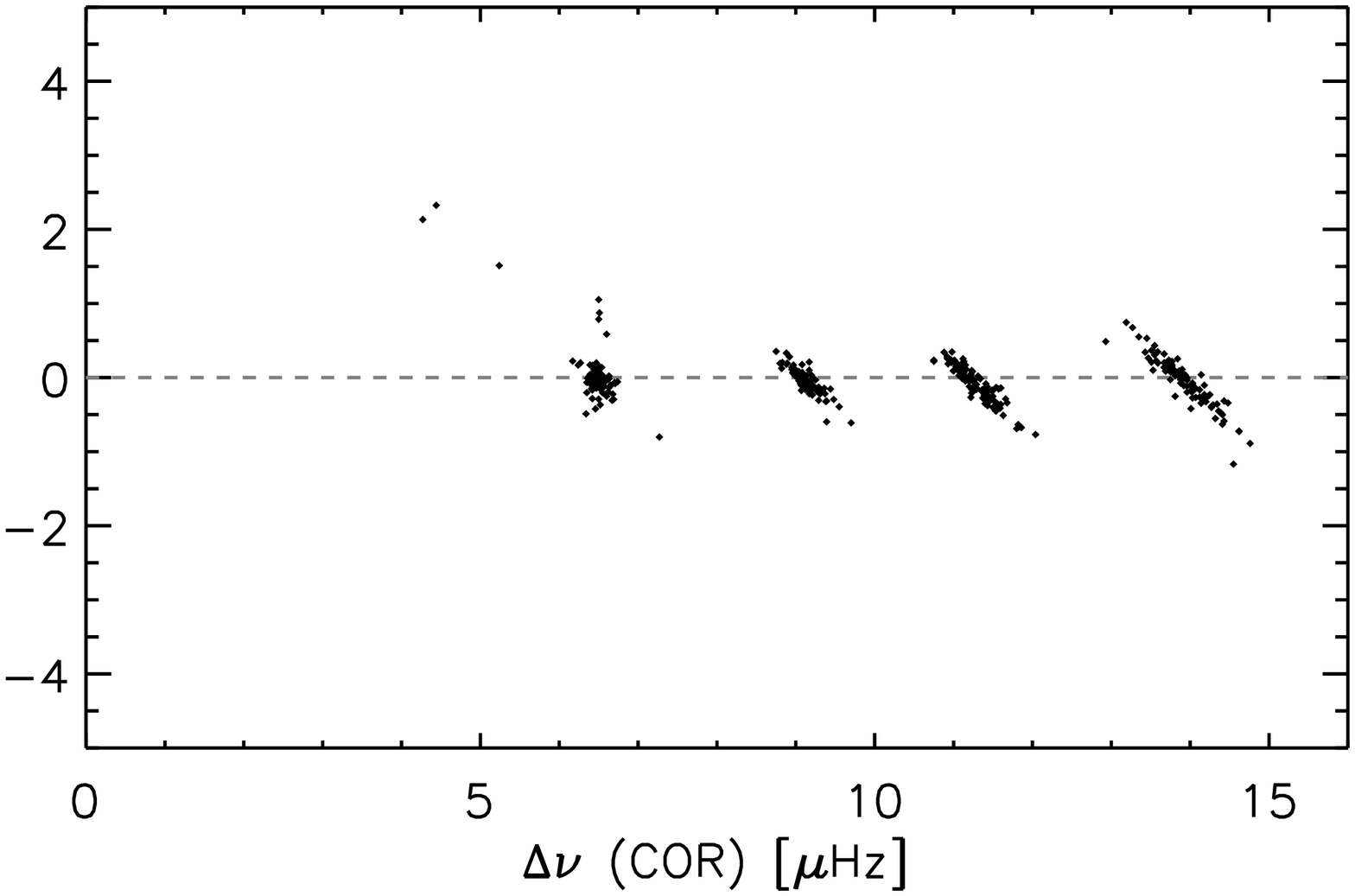}
\end{minipage}
\hfill
\begin{minipage}{5.6cm}
\centering
\includegraphics[width=5.6cm]{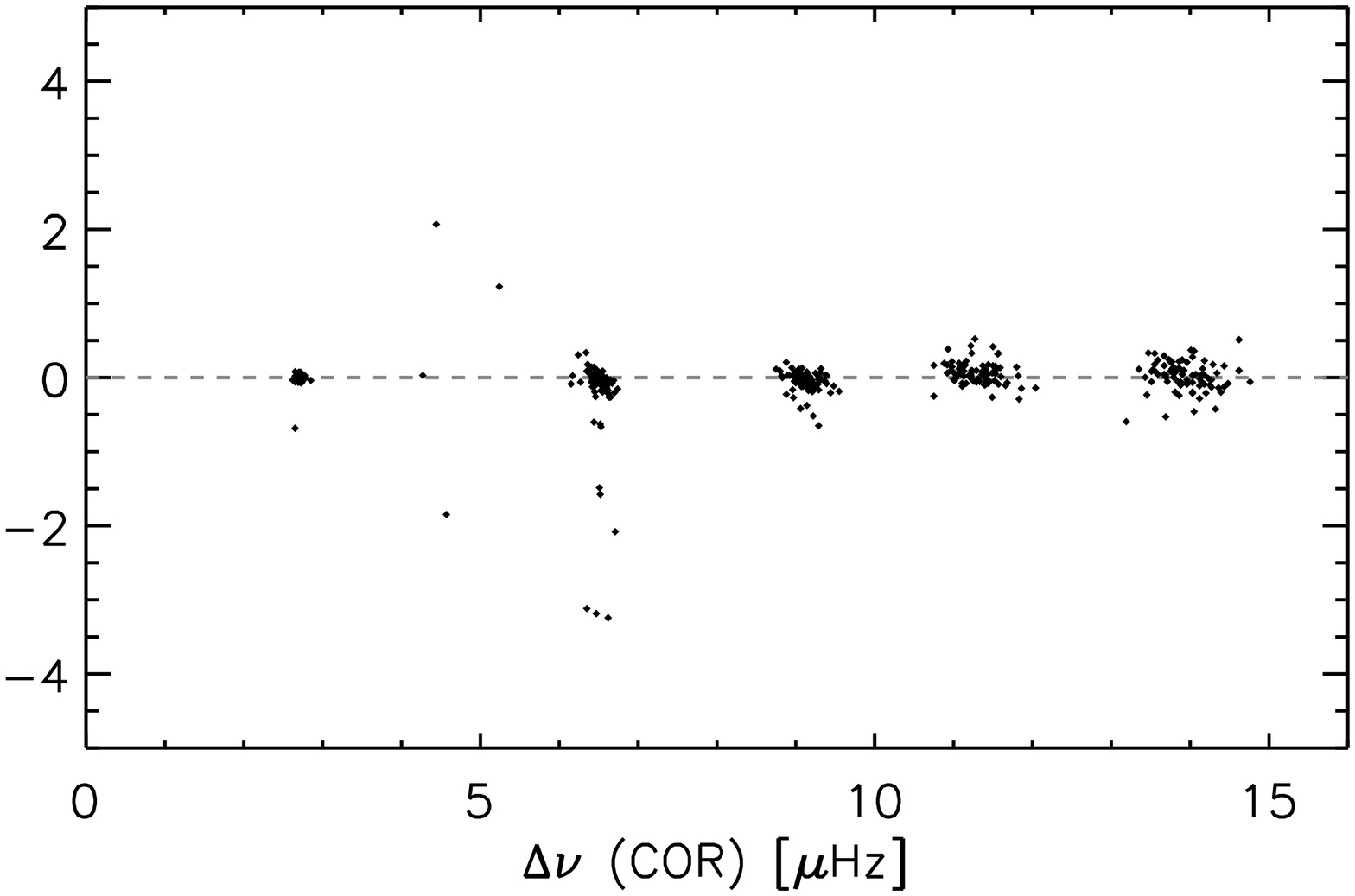}
\end{minipage}
\hfill
\begin{minipage}{5.6cm}
\centering
\includegraphics[width=5.6cm]{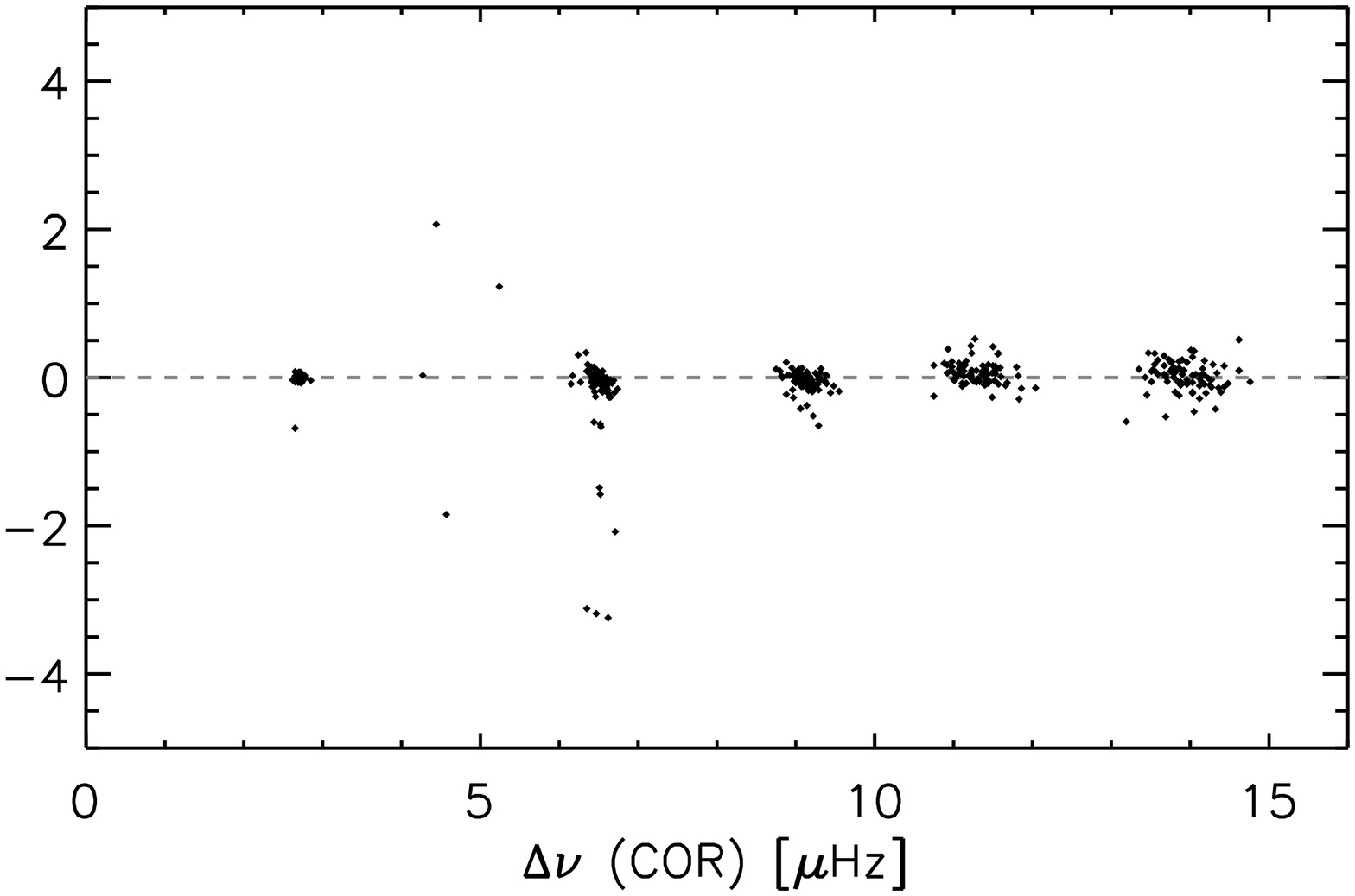}
\end{minipage}
\caption{The differences in the simulation results between different methods, from left to right CAN-COR, OCT II - COR, OCT I - COR are shown as a function of the value returned by COR. The top row is for $\nu_{\rm max}$ and the bottom row for $\Delta \nu$. The gray dashed line indicates agreement between the methods.}
\label{ressim1}
\end{figure*}

\begin{figure*}
\begin{minipage}{5.6cm}
\centering
\includegraphics[width=5.6cm]{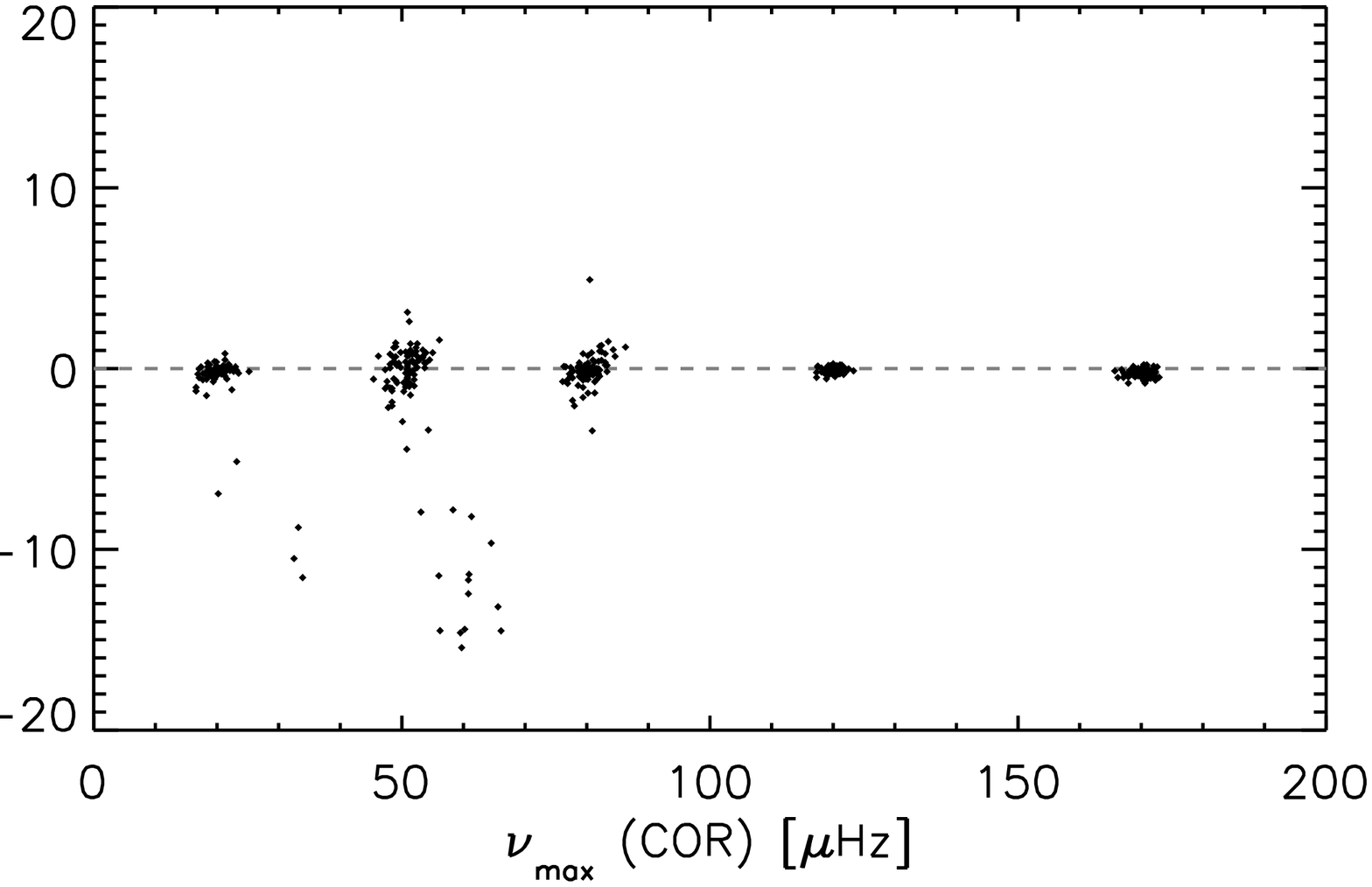}
\end{minipage}
\hfill
\begin{minipage}{5.6cm}
\centering
\includegraphics[width=5.6cm]{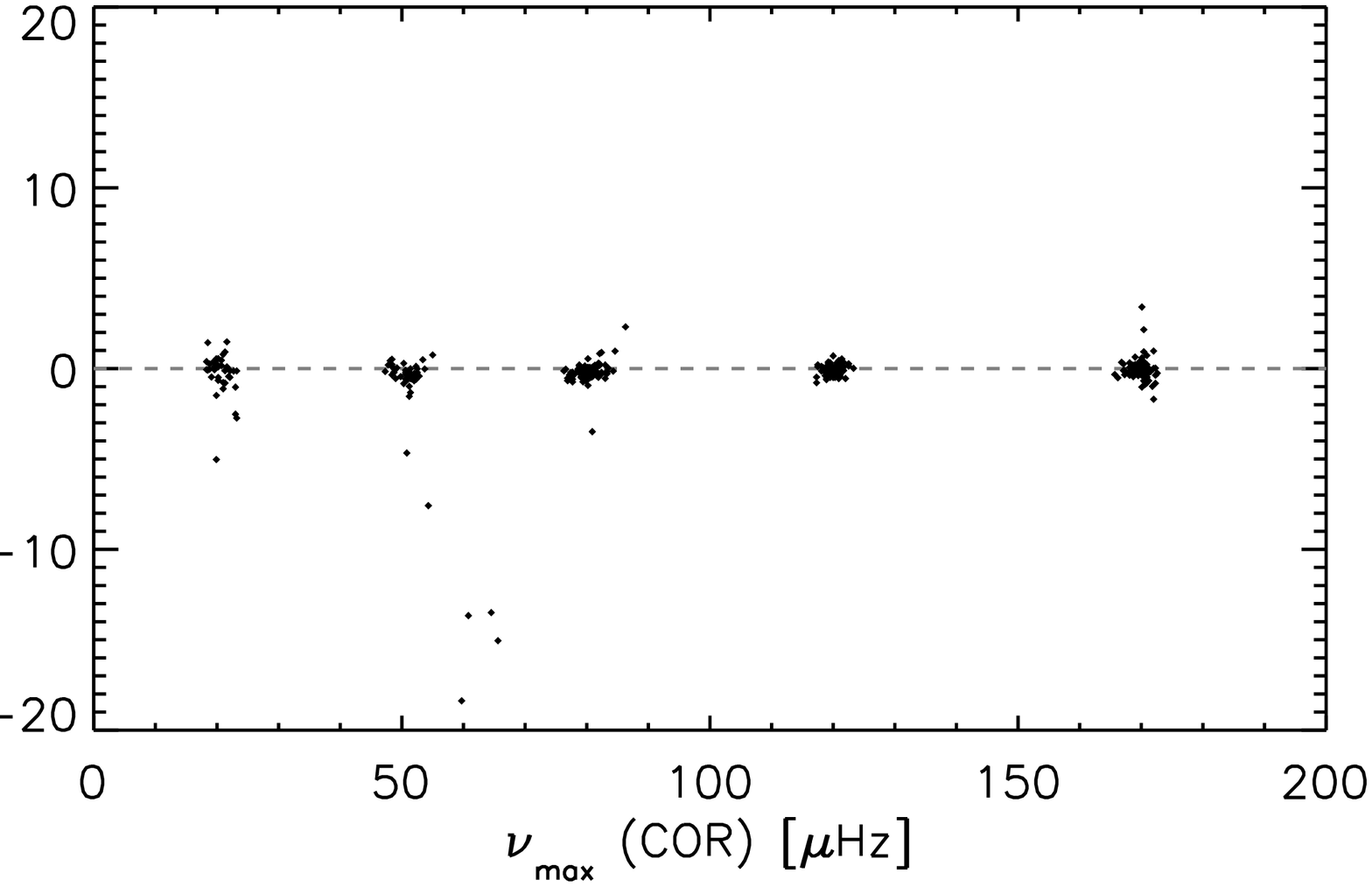}
\end{minipage}
\hfill
\begin{minipage}{5.6cm}
\centering
\includegraphics[width=5.6cm]{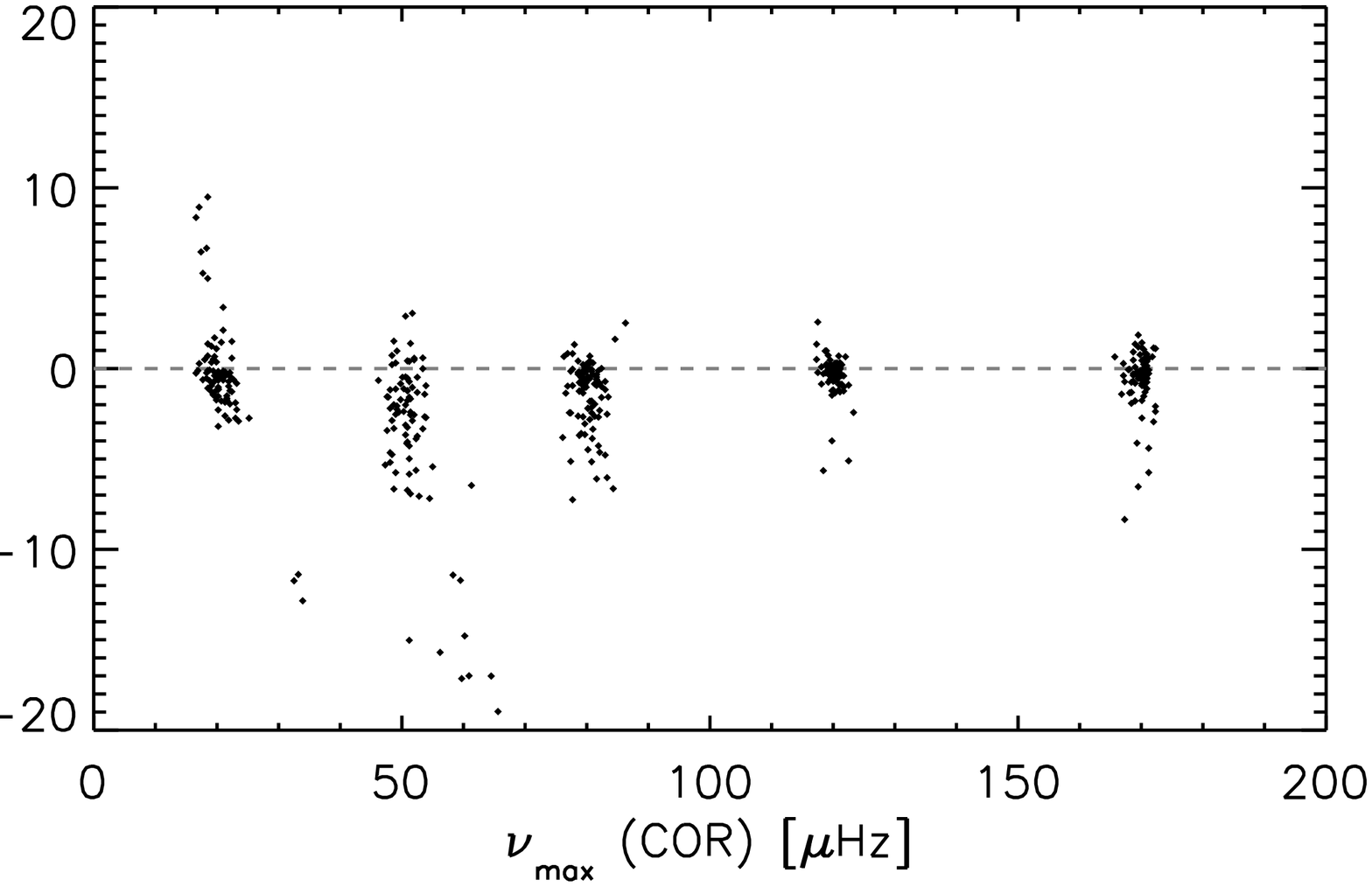}
\end{minipage}
\hfill
\begin{minipage}{5.6cm}
\centering
\includegraphics[width=5.6cm]{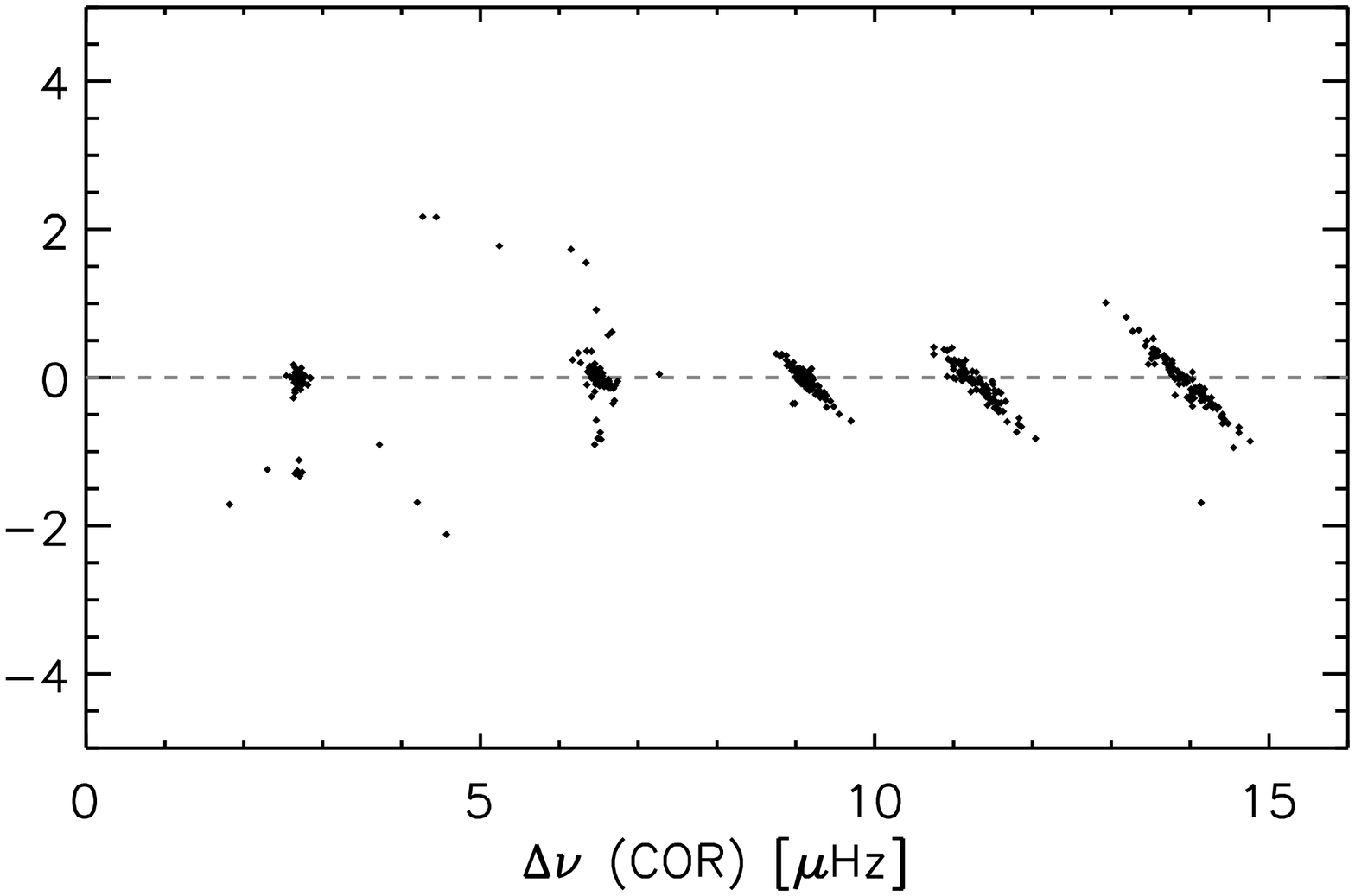}
\end{minipage}
\hfill
\begin{minipage}{5.6cm}
\centering
\includegraphics[width=5.6cm]{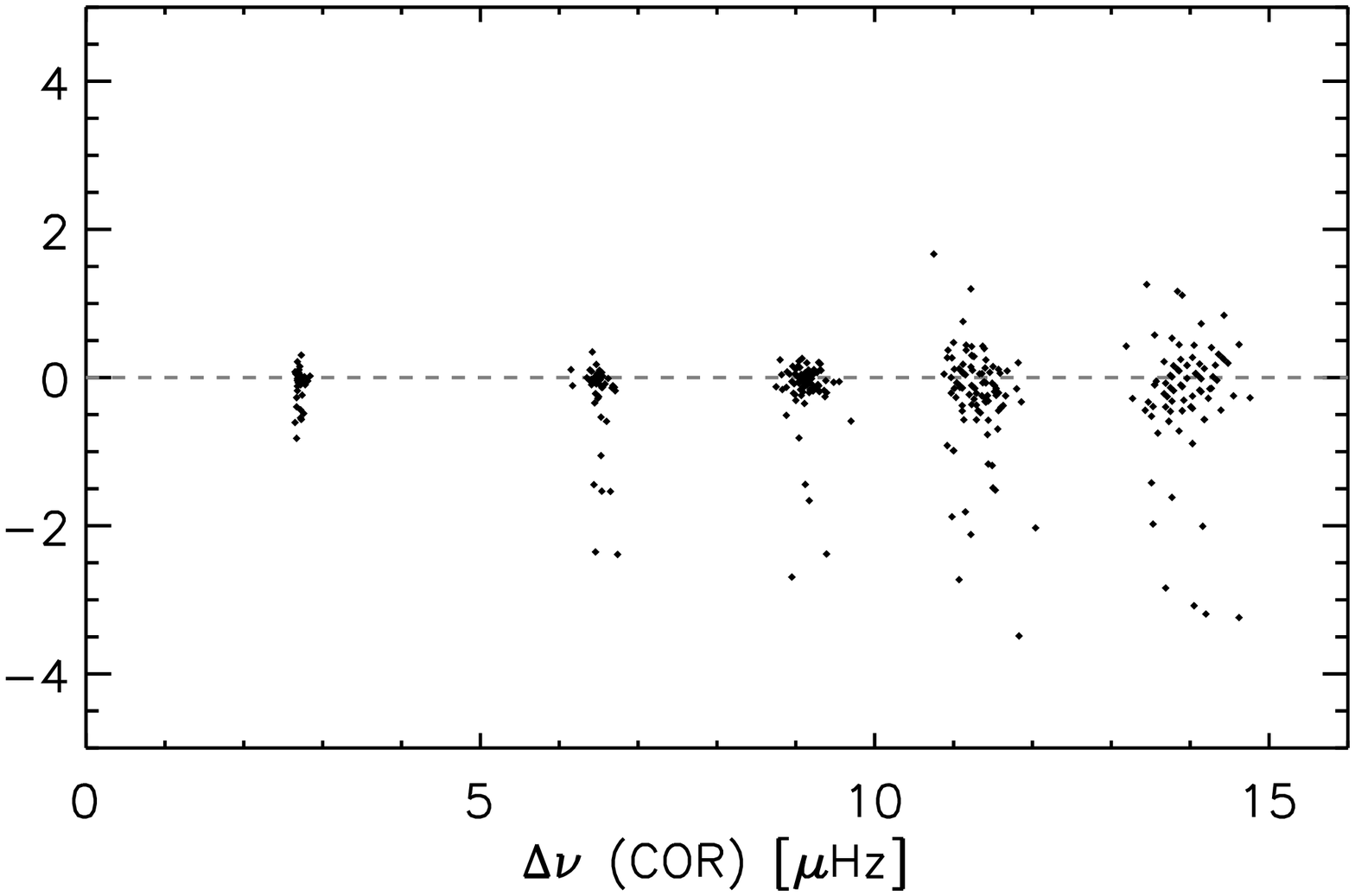}
\end{minipage}
\hfill
\begin{minipage}{5.6cm}
\centering
\includegraphics[width=5.6cm]{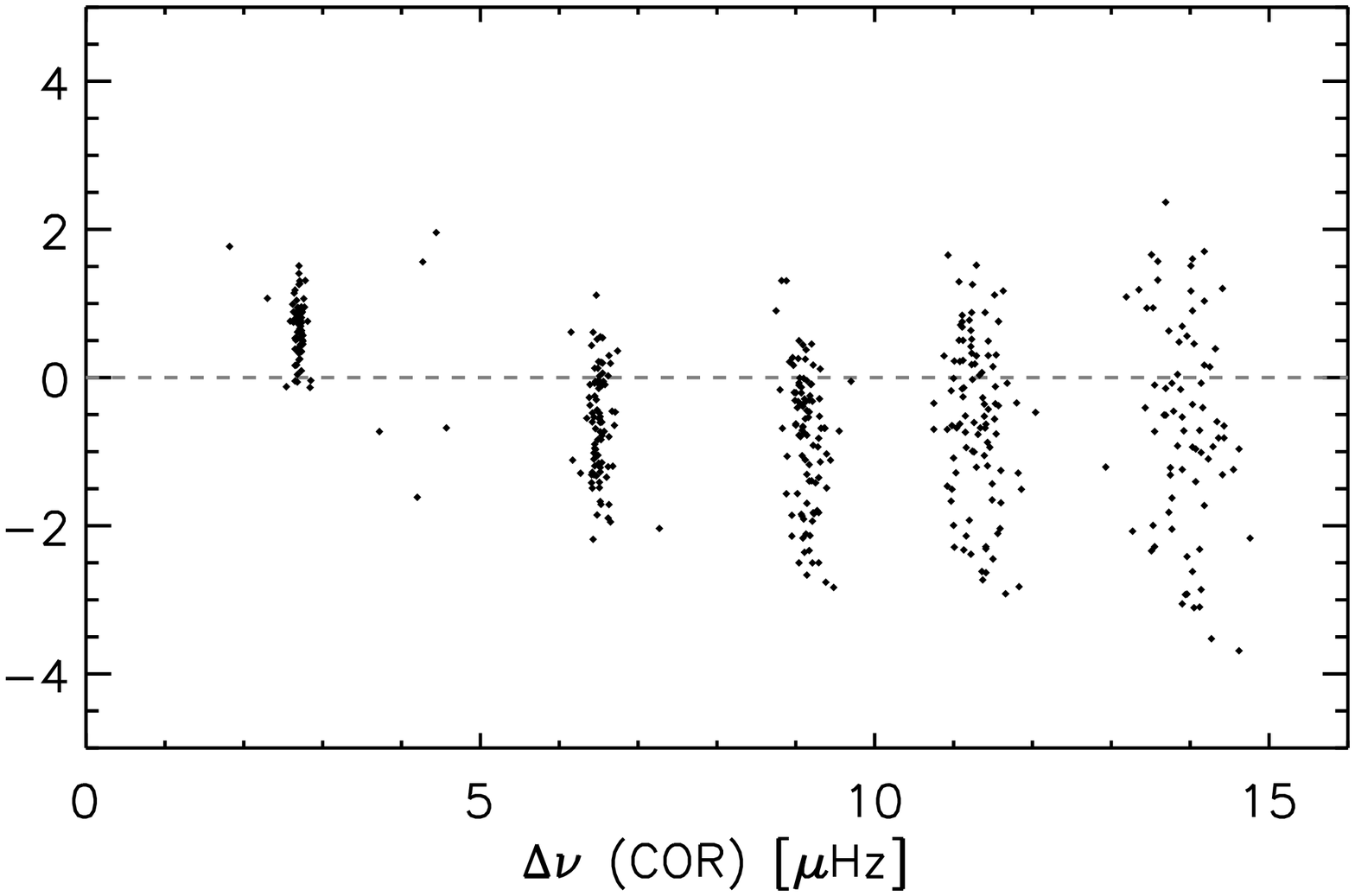}
\end{minipage}
\caption{Same as Fig.~\ref{ressim1} but now from left to right for SYD - COR, A2Z - COR and, DLB - COR}
\label{ressim1a}
\end{figure*}

\begin{figure*}
\begin{minipage}{5.6cm}
\centering
\includegraphics[width=5.5cm]{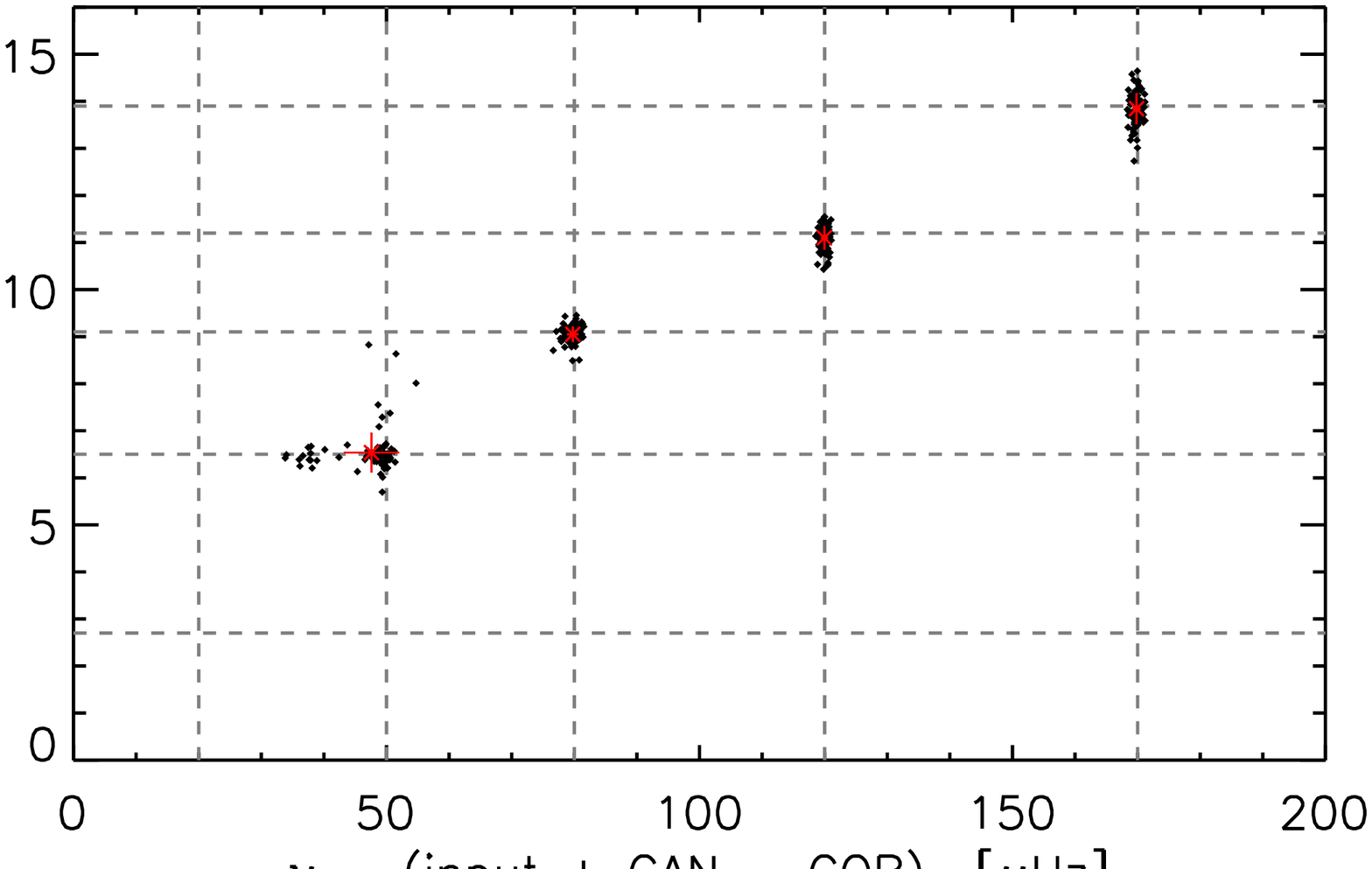}
\end{minipage}
\hfill
\begin{minipage}{5.6cm}
\centering
\includegraphics[width=5.5cm]{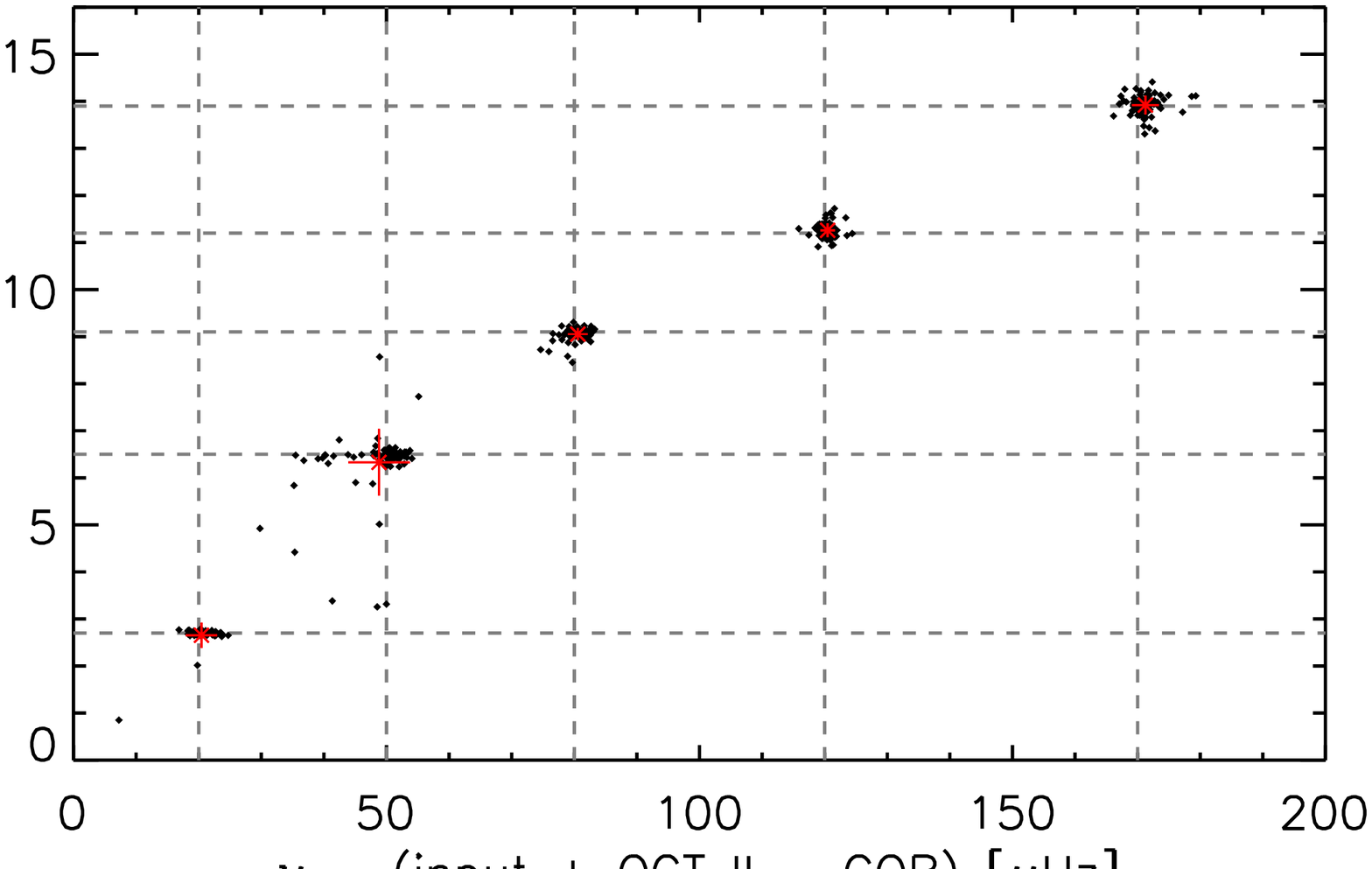}
\end{minipage}
\hfill
\begin{minipage}{5.6cm}
\centering
\includegraphics[width=5.5cm]{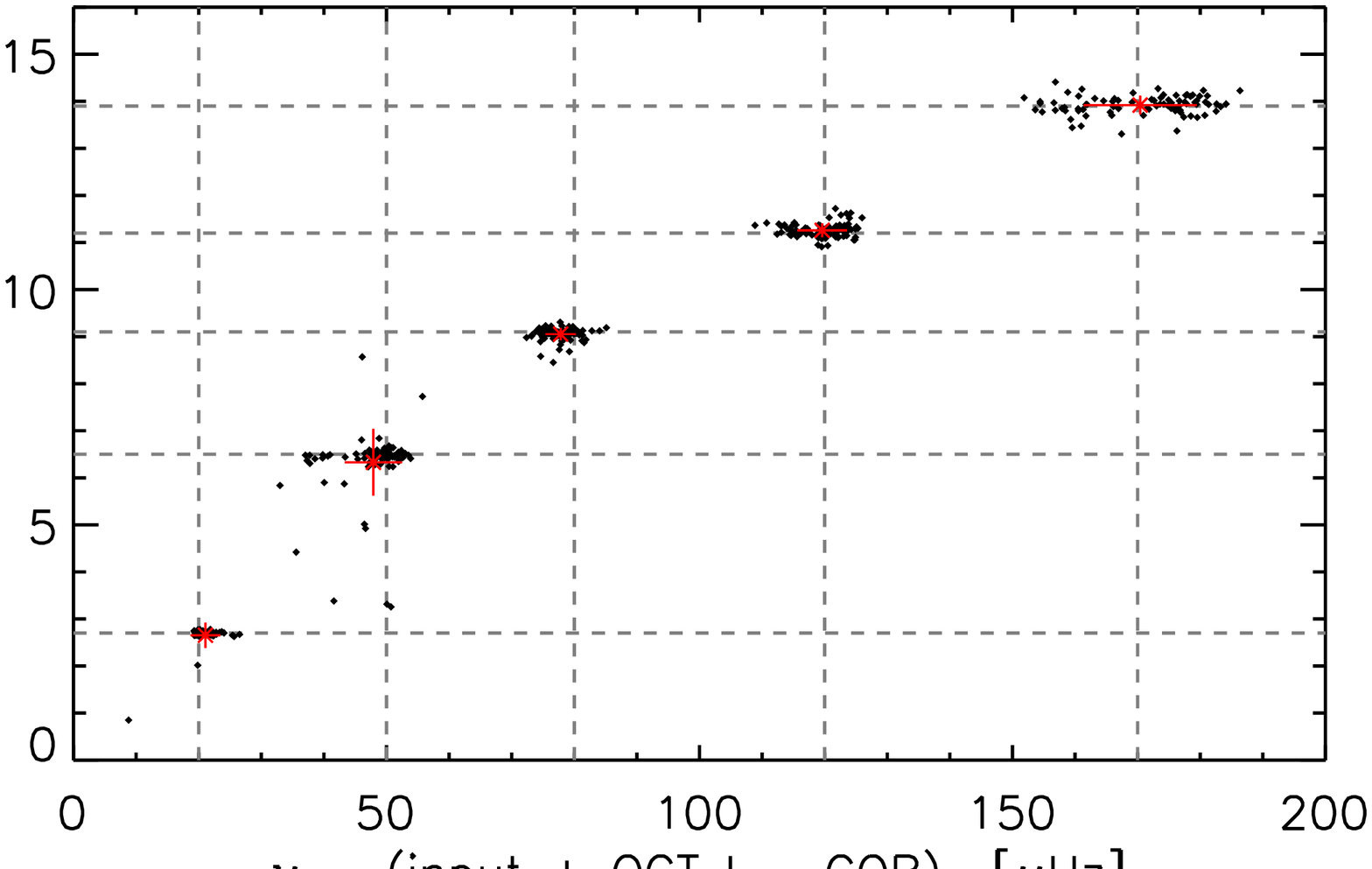}
\end{minipage}
\hfill
\begin{minipage}{5.6cm}
\centering
\includegraphics[width=5.5cm]{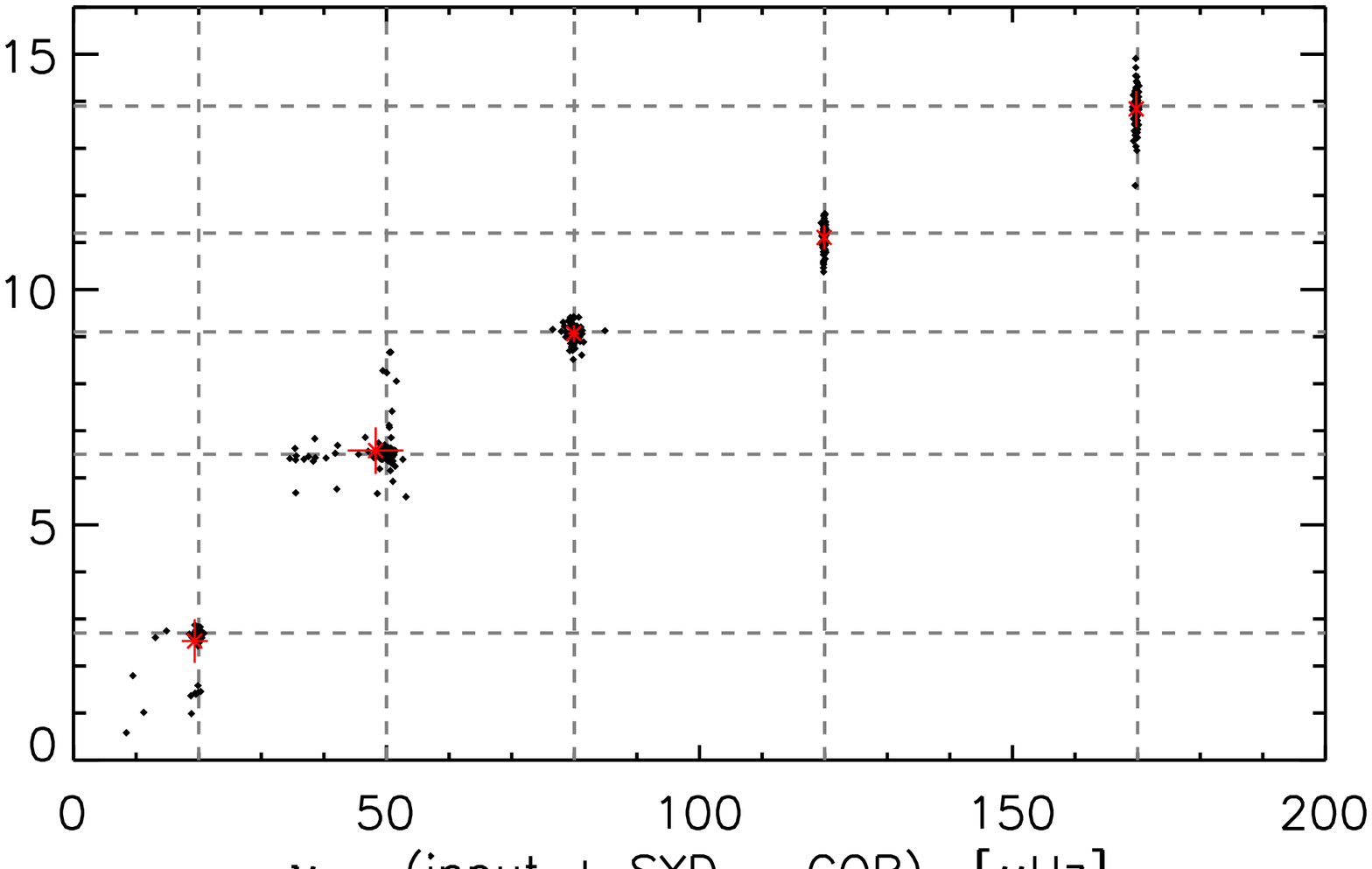}
\end{minipage}
\hfill
\begin{minipage}{5.6cm}
\centering
\includegraphics[width=5.5cm]{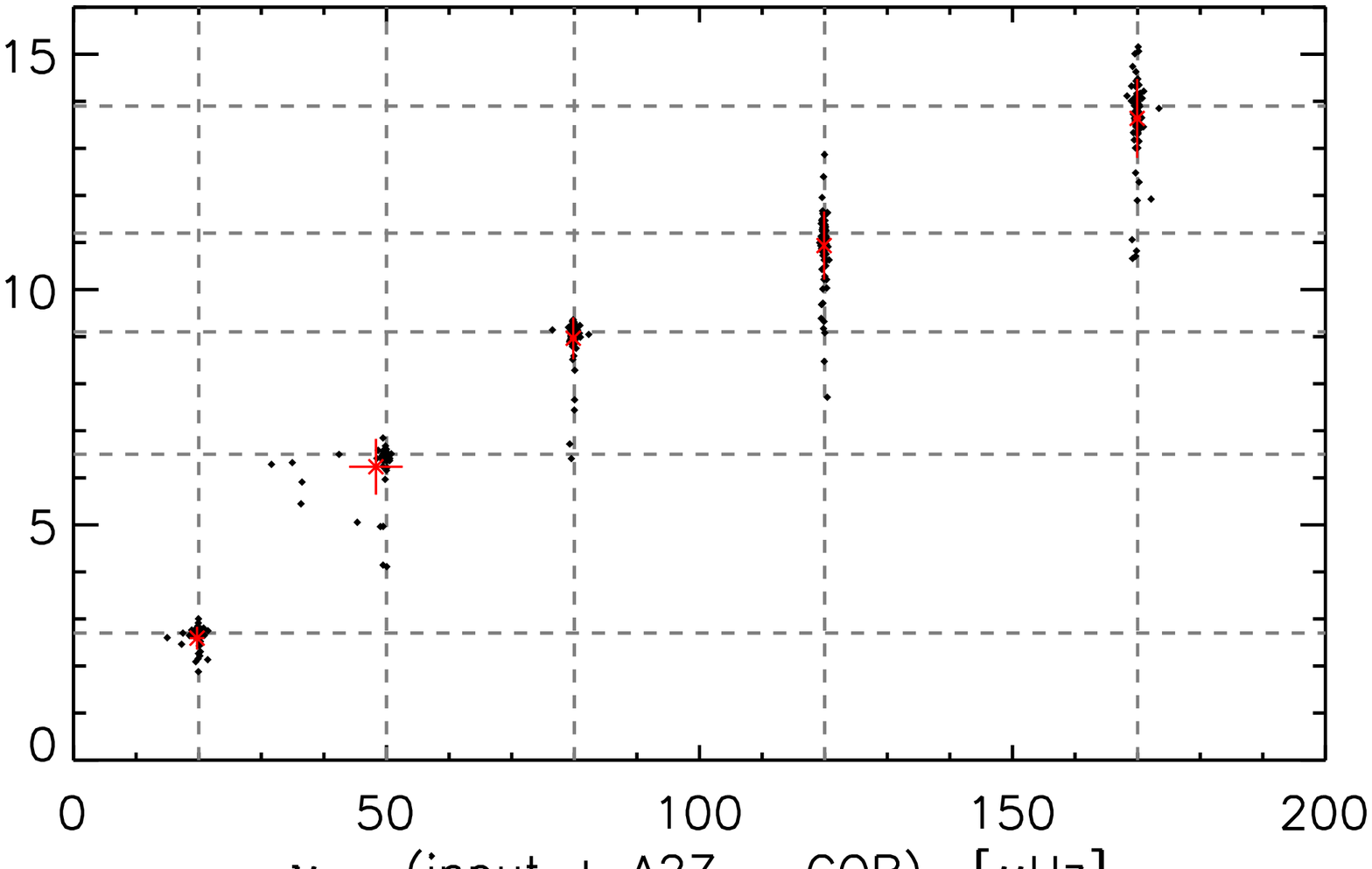}
\end{minipage}
\hfill
\begin{minipage}{5.6cm}
\centering
\includegraphics[width=5.5cm]{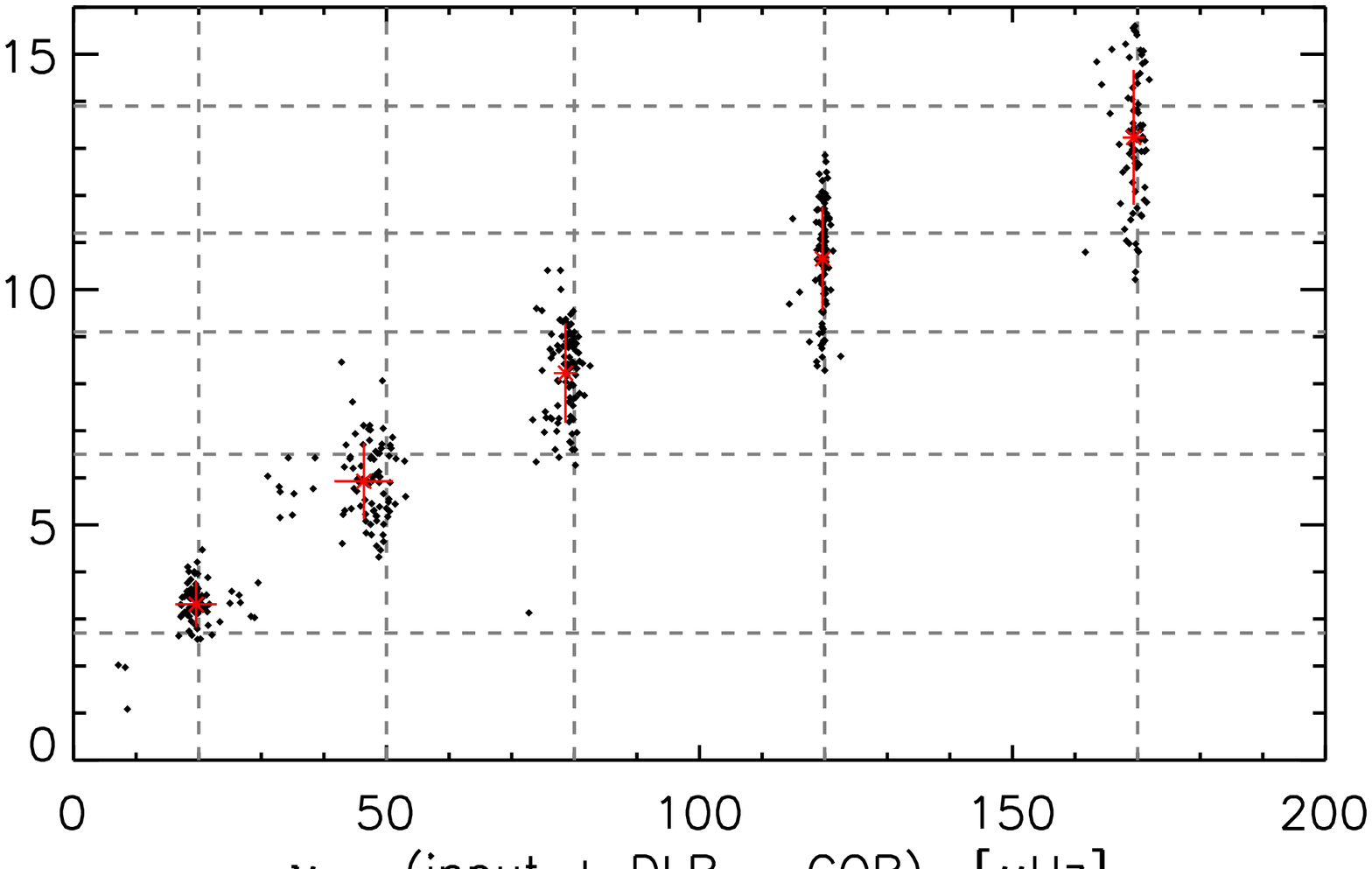}
\end{minipage}
\caption{Same as Fig.~\ref{ressimnoerror}, but here we show the difference in the results from different methods, i.e., from left to right and top to bottom CAN - COR, OCT II - COR, OCT I - COR, SYD - COR, A2Z - COR and, DLB - COR, at the position of the input parameters of the simulation. The mean values and the scatter are indicated with the red asterisks.}
\label{ressim2}
\end{figure*}

\begin{table*}
\begin{minipage}{\linewidth}
\caption{Mean and standard deviations of the computed $\nu_{\rm max}$ for synthetic timeseries. }
\label{outputnumax}
\centering
\begin{tabular}{ccccccccc}
\hline\hline
sim & $\nu_{\rm max}$ & $\nu_{\rm max}$ & $\nu_{\rm max}$ & $\nu_{\rm max}$  & $\nu_{\rm max}$ & $\nu_{\rm max}$ & $\nu_{\rm max}$ & $\nu_{\rm max}$\\
 & input & CAN & COR & OCT I & OCT II & SYD & A2Z & DLB \\
 & $\mu$Hz  & $\mu$Hz & $\mu$Hz & $\mu$Hz & $\mu$Hz & $\mu$Hz & $\mu$Hz & $\mu$Hz\\
\hline
1 & 20.0 & 19.6 $\pm$ 1.0 & 20.4 $\pm$ 2.3 & 21.8 $\pm$ 1.2 & 21.2 $\pm$ 1.2 & 19.9 $\pm$ 2.0 & 20.1 $\pm$ 1.8 & 20.1 $\pm$ 2.1 \\
6 & 20.0 & 19.4 $\pm$ 1.1 & 20.3 $\pm$ 2.3 & 21.9 $\pm$ 1.2 & 21.1 $\pm$ 1.7 & 19.5 $\pm$ 2.4 & 20.6 $\pm$ 1.9 & 20.3 $\pm$ 2.3 \\
2 & 50.0 & 49.6 $\pm$ 2.0 & 52.1 $\pm$ 4.0 & 50.0 $\pm$ 2.3 & 50.8 $\pm$ 2.9 & 50.1 $\pm$ 3.6 & 50.0 $\pm$ 2.3 & 48.0 $\pm$ 3.3 \\
7 & 50.0 & 49.6 $\pm$ 2.3 & 52.9 $\pm$ 4.4 & 50.0 $\pm$ 2.7 & 51.1 $\pm$ 2.6 & 50.1 $\pm$ 3.6 & 49.5 $\pm$ 3.3 & 47.6 $\pm$ 4.8 \\
3 & 80.0 & 79.9 $\pm$ 1.4 & 80.2 $\pm$ 1.9 & 78.0 $\pm$ 3.6 & 80.8 $\pm$ 1.7 & 80.1 $\pm$ 2.3 & 80.1 $\pm$ 2.1 & 78.8 $\pm$ 2.5 \\
8 & 80.0 & 79.8 $\pm$ 1.4 & 79.9 $\pm$ 1.8 & 77.9 $\pm$ 3.5 & 80.7 $\pm$ 1.5 & 79.8 $\pm$ 2.3 & 79.8 $\pm$ 1.8 & 78.7 $\pm$ 2.2 \\
4 & 120.0 & 120.1 $\pm$ 1.0 & 120.1 $\pm$ 1.1 & 119.8 $\pm$ 3.8 & 120.6 $\pm$ 1.4 & 120.0 $\pm$ 1.1 & 120.0 $\pm$ 1.2 & 119.8 $\pm$ 1.3 \\
9 & 120.0 & 119.8 $\pm$ 1.0 & 119.8 $\pm$ 1.2 & 118.9 $\pm$ 4.1 & 120.4 $\pm$ 1.4 & 119.7 $\pm$ 1.3 & 119.7 $\pm$ 1.3 & 119.4 $\pm$ 1.6 \\
5 & 170.0 & 169.8 $\pm$ 1.5 &170.0 $\pm$ 1.4 & 170.4 $\pm$ 9.2 & 171.2 $\pm$ 2.6 & 169.7 $\pm$ 1.4 & 169.8 $\pm$ 0.9 & 169.2 $\pm$ 2.2 \\
10 & 170.0 & 169.7 $\pm$ 1.5 &169.9 $\pm$ 1.8 & 169.2 $\pm$ 9.9 & 170.8 $\pm$ 3.4 & 169.7 $\pm$ 1.9 & 169.8 $\pm$ 0.9 & 168.5 $\pm$ 2.9 \\
\hline
\end{tabular}
\end{minipage}
\end{table*}

\begin{table*}
\begin{minipage}{\linewidth}
\caption{Mean and standard deviations of the computed $\meandnu$ for synthetic timeseries. }
\label{outputdnu}
\centering
\begin{tabular}{cccccccc}
\hline\hline
sim & $\Delta \nu$ & $\Delta \nu$ & $\Delta \nu$  & $\Delta \nu$ & $\Delta \nu$ & $\Delta \nu$ & $\Delta \nu$\\
 & input & CAN & COR & OCT & SYD & A2Z & DLB\\
 & $\mu$Hz & $\mu$Hz  & $\mu$Hz & $\mu$Hz & $\mu$Hz & $\mu$Hz & $\mu$Hz \\
\hline
1 & 2.7 &  & 2.7 $\pm$ 0.2 & 2.7 $\pm$ 0.2 & 2.6 $\pm$ 0.5 & 2.6 $\pm$ 0.2 & 3.3 $\pm$ 0.4\\
6 & 2.7 &  & 2.70 $\pm$ 0.04 & 2.7 $\pm$ 0.2 & 2.7 $\pm$ 0.2 & 2.7 $\pm$ 0.2 & 3.4 $\pm$ 1.3\\
2 & 6.5 & 6.5 $\pm$ 0.3 & 6.4 $\pm$ 0.3 & 6.3 $\pm$ 0.7 & 6.5 $\pm$ 0.3 & 6.2 $\pm$ 0.6 & 5.9 $\pm$ 0.7\\
7 & 6.5 & 6.5 $\pm$ 0.2 & 6.5 $\pm$ 0.1 & 6.2 $\pm$ 0.7 & 6.5 $\pm$ 0.3 & 6.0 $\pm$ 0.8 & 6.1 $\pm$ 0.9\\
3 & 9.1 & 9.1 $\pm$ 0.1 & 9.1 $\pm$ 0.2 & 9.1 $\pm$ 0.2 & 9.1 $\pm$ 0.1 & 9.0 $\pm$ 0.5 & 8.3 $\pm$ 1.0\\
8 & 9.1 & 9.1 $\pm$ 0.1 & 9.1 $\pm$ 0.2 & 9.1 $\pm$ 0.2 & 9.1 $\pm$ 0.1 & 9.0 $\pm$ 0.6 & 8.4 $\pm$ 0.9\\
4 & 11.2 & 11.2 $\pm$ 0.1 & 11.3 $\pm$ 0.3 & 11.4 $\pm$ 0.3 & 11.2 $\pm$ 0.1 & 11.0 $\pm$ 0.7 & 10.7 $\pm$ 1.1\\
9 & 11.2 & 11.2 $\pm$ 0.1 & 11.3 $\pm$ 0.3 & 11.4 $\pm$ 0.3 & 11.2 $\pm$ 0.1 & 11.0 $\pm$ 0.8 & 10.9 $\pm$ 1.1\\
5 & 13.9 & 13.9 $\pm$ 0.1 & 13.9 $\pm$ 0.3 & 14.0 $\pm$ 0.4 & 13.9 $\pm$ 0.2 & 13.7 $\pm$ 0.9 & 13.3 $\pm$ 1.4\\
10 & 13.9 & 13.9 $\pm$ 0.1 &14.0 $\pm$ 0.3 & 14.1 $\pm$ 0.3 & 13.9 $\pm$ 0.1 & 13.8 $\pm$ 0.9 & 13.7 $\pm$ 1.5\\
\hline
\end{tabular}
\end{minipage}
\end{table*}

\begin{table*}
\begin{minipage}{\linewidth}
\caption{The standard deviations of the differences in results for $\nu_{\rm max}$  from different methods.}
\label{outputdifnumax}
\centering
\begin{tabular}{cccccccc}
\hline\hline
sim & $\nu_{\rm max}$ & $\sigma_{\nu_{\rm max}}$ & $\sigma_{\nu_{\rm max}}$ & $\sigma_{\nu_{\rm max}}$ & $\sigma_{\nu_{\rm max}}$ & $\sigma_{\nu_{\rm max}}$ & $\sigma_{\nu_{\rm max}}$ \\
 & \tiny{input} & \tiny{CAN-COR} &\tiny{OCT I-COR} & \tiny{OCT II-COR} & \tiny{SYD- COR} & \tiny{A2Z-COR} &\tiny{DLB-COR} \\
 & $\mu$Hz  & $\mu$Hz & $\mu$Hz & $\mu$Hz & $\mu$Hz & $\mu$Hz & $\mu$Hz\\
\hline
1 & 20.0 & 2.9 & 2.4 & 2.5 & 2.1& 1.1 & 3.3 \\
6 & 20.0 & 2.3 & 2.5 & 2.7 & 2.5 & 2.1 & 2.7 \\
2 & 50.0 & 4.4 & 4.5 & 4.9 & 4.4 & 4.2 & 4.7 \\
7 & 50.0 & 5.4 & 5.0 & 5.3 & 6.2 & 5.5 & 6.7 \\
3 & 80.0 & 0.9 & 2.4 & 1.6 & 0.9 & 0.5 & 1.9 \\
8 & 80.0 &  0.9 & 2.6 & 1.2 & 0.8 & 0.4 & 1.6 \\
4 & 120.0 & 0.6 & 3.9 & 1.1 & 0.2 & 0.3 & 1.0 \\
9 & 120.0 & 0.7 & 4.2 & 1.3 & 0.8 & 0.3 & 1.2 \\
5 & 170.0 & 0.6 & 9.0 & 2.1 & 0.2 & 0.6 & 1.7 \\
10 & 170.0 & 0.7 & 9.3 & 2.9 & 0.3 & 0.4 & 1.5 \\
\hline
\end{tabular}
\end{minipage}
\end{table*}

\begin{table*}
\begin{minipage}{\linewidth}
\caption{The standard deviations of the differences in results for $\meandnu$ from different methods.}
\label{outputdifdnu}
\centering
\begin{tabular}{ccccccc}
\hline\hline
sim & $\Delta \nu$ &  $\sigma_{\Delta \nu}$ & $\sigma_{\Delta \nu}$ & $\sigma_{\Delta \nu}$ & $\sigma_{\Delta \nu}$\\
 & \tiny{input} & \tiny{CAN-COR} & \tiny{OCT-COR} & \tiny{SYD-COR} & \tiny{A2Z-COR} & \tiny{DLB-COR}\\
 & $\mu$Hz & $\mu$Hz & $\mu$Hz & $\mu$Hz & $\mu$Hz & $\mu$Hz \\
\hline
1 & 2.7 & & 0.3 & 0.5 & 0.2 & 0.5\\
6 & 2.7 & & 0.2 & 0.2 & 0.2 & 1.3\\
2 & 6.5  & 0.4 & 0.7 & 0.5 & 0.6 & 0.8\\
7 & 6.5  & 0.2 & 0.7 & 0.3 & 0.8 & 1.0\\
3 &  9.1 & 0.2 & 0.1 & 0.2 & 0.4 & 1.1\\
8 &  9.1 & 0.2 & 0.1 & 0.2 & 0.6 & 0.9\\
4 & 11.2 & 0.3 & 0.1 & 0.3 & 0.7 & 1.1\\
9 & 11.2 & 0.3 & 0.2 & 0.3 & 0.8 & 1.1\\
5 & 13.9 & 0.3 & 0.2 & 0.4 & 0.8 & 1.4\\
10 & 13.9 & 0.3 & 0.3 & 0.3 & 0.8 & 1.5\\
\hline
\end{tabular}
\end{minipage}
\end{table*}

These synthetic timeseries were analysed using the methods described in the previous section. The results for simulations 1-5  are shown in Fig.~\ref{ressimnoerror} and compared in Figs.~\ref{ressim1}, \ref{ressim1a} and \ref{ressim2}, in which the comparison is shown as a function of the results obtained with method COR. Apart from results for some individual realizations around $\nu_{\rm max}$ of 50 $\mu$Hz, for which the oscillations have the smallest height compared to the background, the results are consistent with the input values, but with a non-negligible scatter. This scatter is inherent to the observations of the star, although not fully independent of the analysis method, and should be taken into account when detailed modelling is performed for a particular star. The mean values of the results of all realizations with the same input parameters and the standard deviations of the results are listed in Tables~\ref{outputnumax} and \ref{outputdnu}.  The results for simulations 6-10 with longer mode life times are very similar to the results obtained for simulations 1-5. Therefore, we conclude that the mode life time does not significantly influence the obtained $\nu_{\rm max}$ and $\meandnu$ at least for the values investigated here.

Figs.~\ref{ressimnoerror}, \ref{ressim1} and \ref{ressim1a} show some interesting differences in the strengths and weaknesses of the different methods. Firstly, from Fig.~\ref{ressimnoerror} and Table~\ref{outputnumax} and \ref{outputdnu} it is clear that all methods provide in general results consistent with the input parameters. Only the Lorentzian fitting from CAN did not return values for $\meandnu$ for the simulations with $\nu_{\rm max}$ at 20 $\mu$Hz, because the frequency peaks were spaced too closely relative to the timebase of the dataset. In general, for the realizations at $\nu_{\rm max}$ of 50 $\mu$Hz and 170 $\mu$Hz the spread in the results is larger due to the relative weakness of the oscillations, and the results of some methods show more spread in $\nu_{\rm max}$ (OCT I), while for others the spread in $\meandnu$ is somewhat larger (A2Z and DLB). 
Secondly, when comparing the results of the different methods with respect to COR (see Figs.~\ref{ressim1} and \ref{ressim1a}) this shows that, apart from the `outliers', the results for $\Delta \nu$ of the different methods agree to sub-$\mu$Hz level. The agreement in $\nu_{\rm max}$ is typically a few $\mu$Hz. Larger differences are only present for oscillations at the highest frequency when using the OCT I method (centroid of Gaussian fit to smoothed power excess).
From the trends in the centre top panel in Fig.~\ref{ressim1} it is clear that for $\nu_{\rm max}$ the spread in the results at low frequencies is largest for results obtained with COR, i.e., the spread of the results on the x-axis (COR) is equivalent to the spread on the y-axis (OCT II-COR). At higher frequencies the differences in $\nu_{\rm max}$ seem to be mainly due to OCT II. In this case the spread in results on the x-axis (COR) is less than the spread on the y-axis (OCT II-COR). Similar reasoning reveals that for $\Delta \nu$ at $\nu_{\rm max} >$ 80 $\mu$Hz the spread in the results obtained with method COR dominate the total spread in the difference between the results CAN - COR (left bottom panel in Fig.~\ref{ressim1}) and SYD - COR (left bottom panel in Fig.~\ref{ressim1a}). 

The differences in the results can be due to differences in the methods and due to the realization noise. To investigate this we show the differences in the results added to the input value (see Fig.~\ref{ressim2}). The scatter in the values in this figure can be interpreted as the differential response of the methods, and are listed in Tables~\ref{outputdifnumax} and \ref{outputdifdnu}. From this table it is notable that for $\meandnu$ of oscillations with $\nu_{\rm max}$ at 80 $\mu$Hz or above, the scatter between OCT and COR is less than the scatter in Fig.~\ref{ressimnoerror}. This indicates that in these cases the scatter due to realization noise is at least as important as the scatter in the results due to the different methods. We also see that the scatter in the difference of the results from SYD - COR and A2Z - COR for $\nu_{\rm max}$ is less for $\nu_{\rm max} \ge$ 80 $\mu$Hz. For the other simulations we cannot distinguish between realization noise and uncertainties, possibly due to the effect of beating between signal and background noise, i.e. changes in the amplitude and hence power when the signal and noise are of similar magnitudes due to the fact that in the amplitude spectrum the signal and noise are added as complex numbers and their relative phases can change significantly. These results indicate that the spread in results due to realization noise can be larger than the difference in the results between different methods, but this depends on the frequency range and the methods used.

Finally, the error bars in Fig.~\ref{ressimnoerror} show that uncertainties on the results are in general not a reliable indication of the spread in the results. We discuss the issue of the uncertainties further in Sect 5.

\section{Observations}
For the present investigation we used data obtained by the NASA \textit{Kepler} satellite taken during the 10 day commissioning run (Q0), the one-month first roll (Q1) and the three-month second roll (Q2), all taken in long-cadence mode \citep[29.4-minute sampling, not entirely regular due to changes in the correction to barycentric times caused by the satelite orbit, see for more details][]{jenkins2010}. These data are of unprecedented quality, although there are still some outliers and artefacts present in the timeseries of some stars.
Before the power spectra are computed, outliers present in the timeseries need to be treated. Their positions can either be filled with zeroes or with a mean flux computed from surrounding observations, or left blank. In Q2 there was a safe-mode event in the first month during which the satellite warmed up and the subsequent thermal relaxation produced photometric responses. In addition there were three separate adjustments of the pointing at the $\sim$0.05 pixel level, leading to step functions in the photometry in some cases. 
We investigated several ways to remove these signals, such as removing the affected part of the timeseries, or splitting the light curve into several parts, and correcting them individually. These different methods seem to work equally well and therefore we do not expect significant differences in the results of the oscillation parameters described here due to these corrections. 

\begin{figure*}
\begin{minipage}{5.6cm}
\centering
\includegraphics[width=5.6cm]{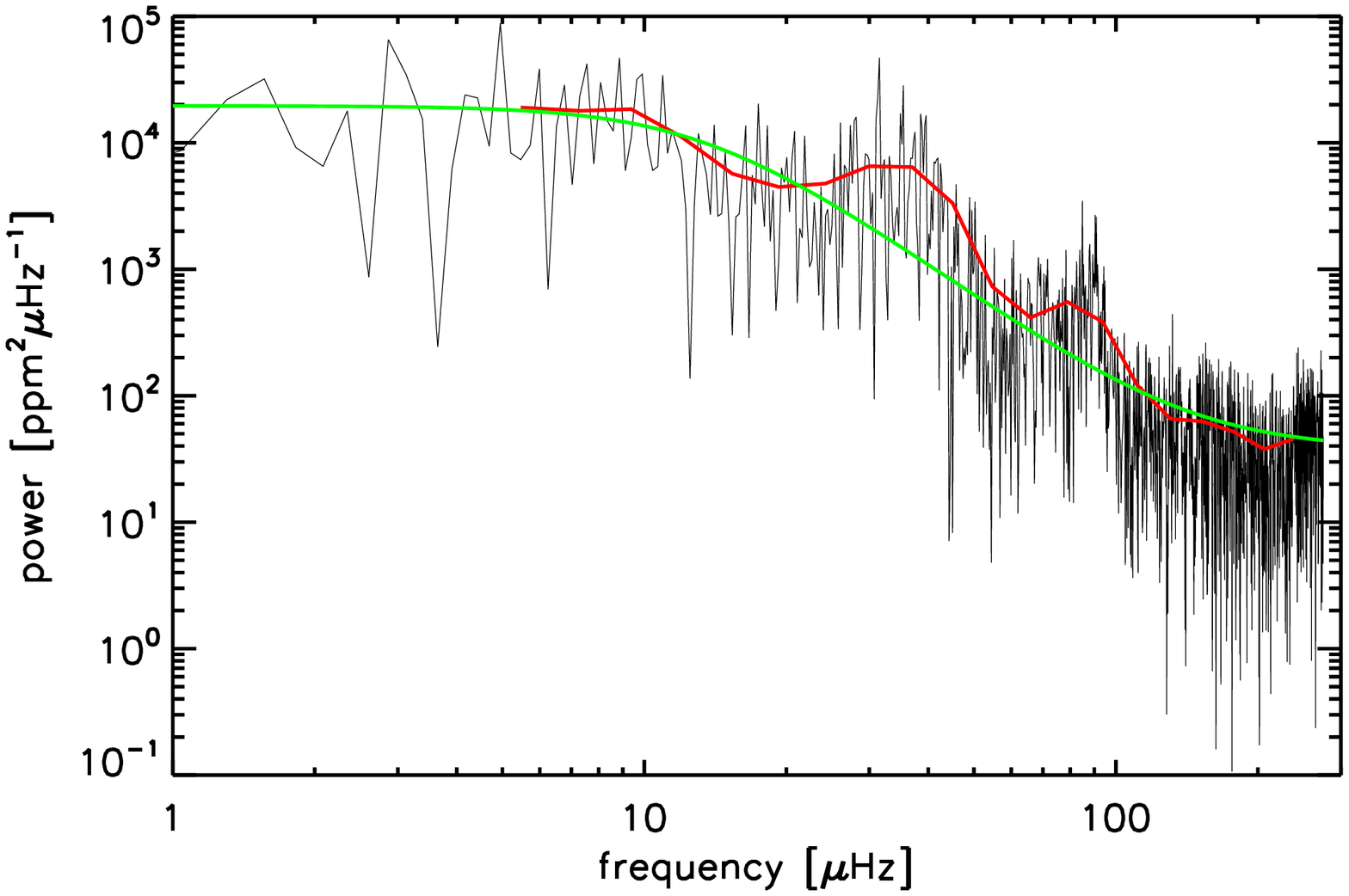}
\end{minipage}
\hfill
\begin{minipage}{5.6cm}
\centering
\includegraphics[width=5.6cm]{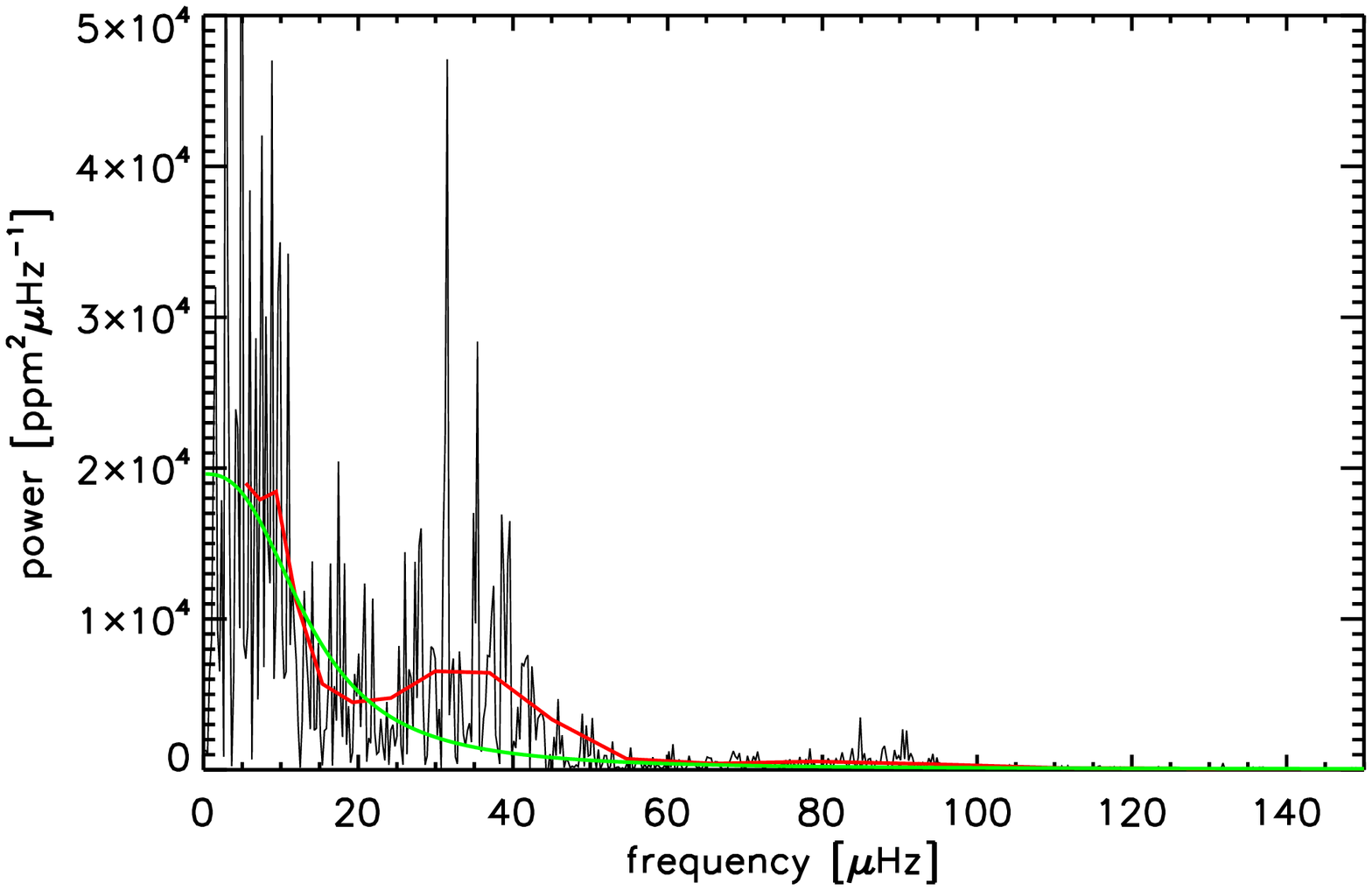}
\end{minipage}
\hfill
\begin{minipage}{5.6cm}
\centering
\includegraphics[width=5.6cm]{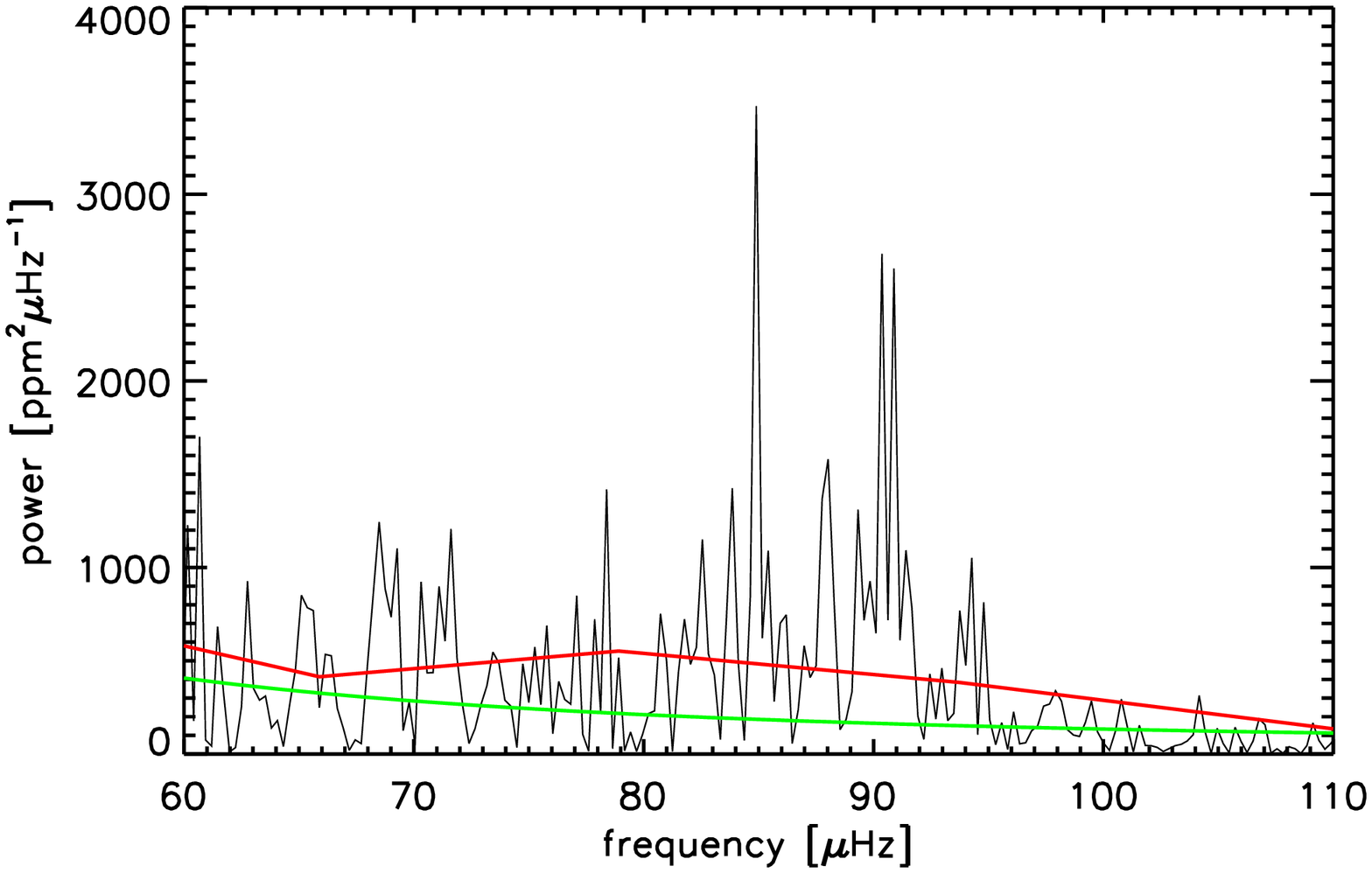}
\end{minipage}
\caption{Power spectra of a target (KIC4813971) on CCD 2 showing power excess due to solar-like oscillations at frequencies around 35 $\mu$Hz and the artefact due to desaturation around 87 $\mu$Hz. The red line indicates a binned power spectrum and the green line a fit to the background.}
\label{art}
\end{figure*}

Investigation of the data taken during the first roll (Q1) revealed that many science targets exhibit non-sinusoidal oscillations with a 3.2hr period (87 $\mu$Hz).  This phenomenon is associated with cycling of a heater on one of the Kepler reaction wheels. 
For an example of this signature see Fig.~\ref{art}. Due to the non-sinusoidal nature of the signal the resulting signature in the power spectrum mimics solar-like oscillations in terms of its width and near equidistant features. We investigated whether it is related to CCD module or position on the CCD module. We found that it seems to be worse for CCD modules located on the edge of the complete 42-CCD mozaic and mostly on one side, i.e. CCDs 2, 3, 4, 6, 10, 11. So far this was only investigated for stars observed during the Q0 and Q1 runs.

\begin{figure*}
\begin{minipage}{\linewidth}
\centering
\includegraphics[width=\linewidth]{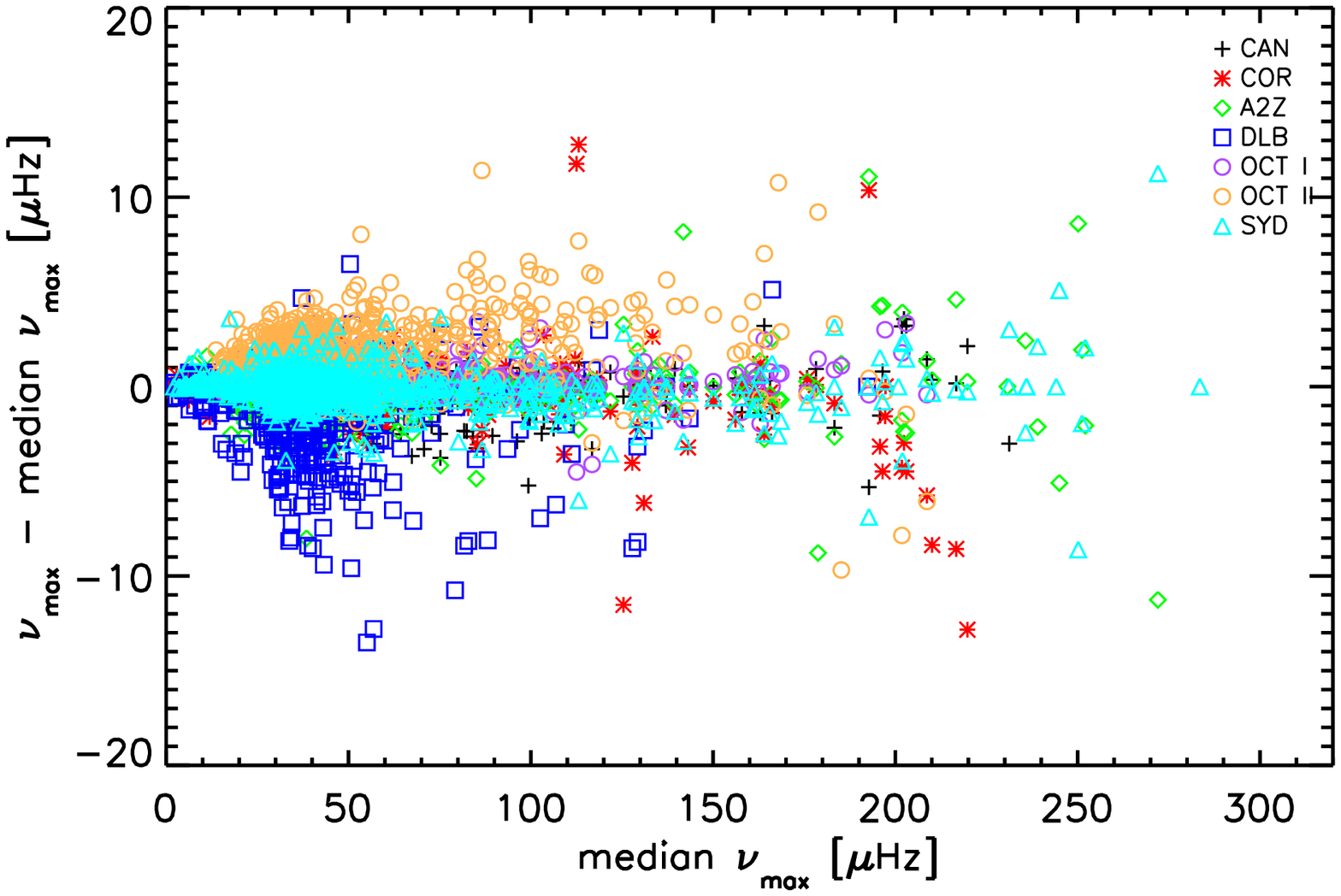}
\end{minipage}
\hfill
\begin{minipage}{\linewidth}
\centering
\includegraphics[width=\linewidth]{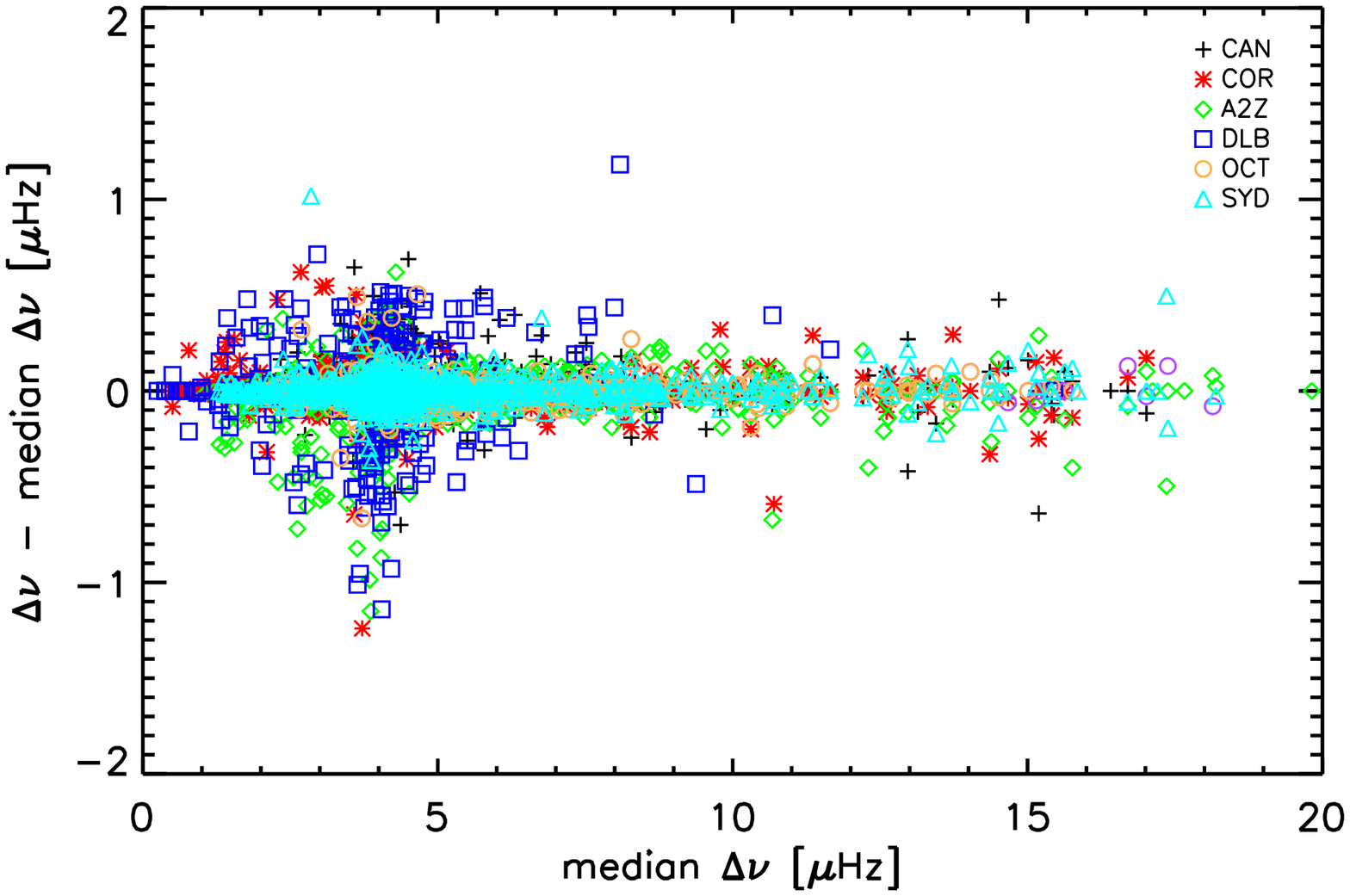}
\end{minipage}
\hfill
\caption{Results of all methods represented as the value - median value versus the median value for $\nu_{\rm max}$ (top) and $\Delta \nu$ (bottom). The different colours and symbols represent results from the different methods, as indicated in the legend.}
\label{figresults}
\end{figure*}

\begin{figure*}
\begin{minipage}{8.4 cm}
\centering
\includegraphics[width=8.4cm]{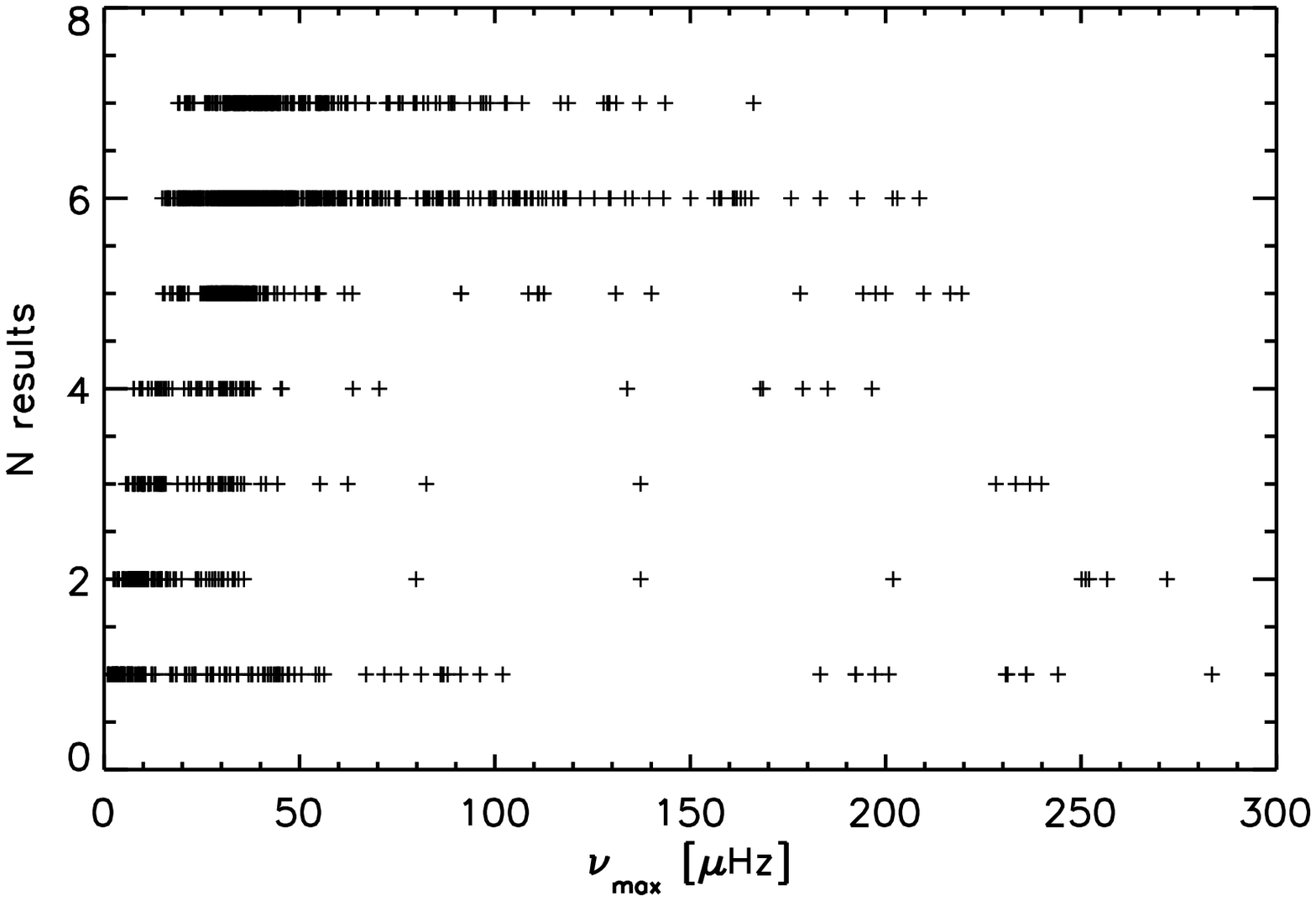}
\end{minipage}
\hfill
\begin{minipage}{8.4cm}
\centering
\includegraphics[width=8.4cm]{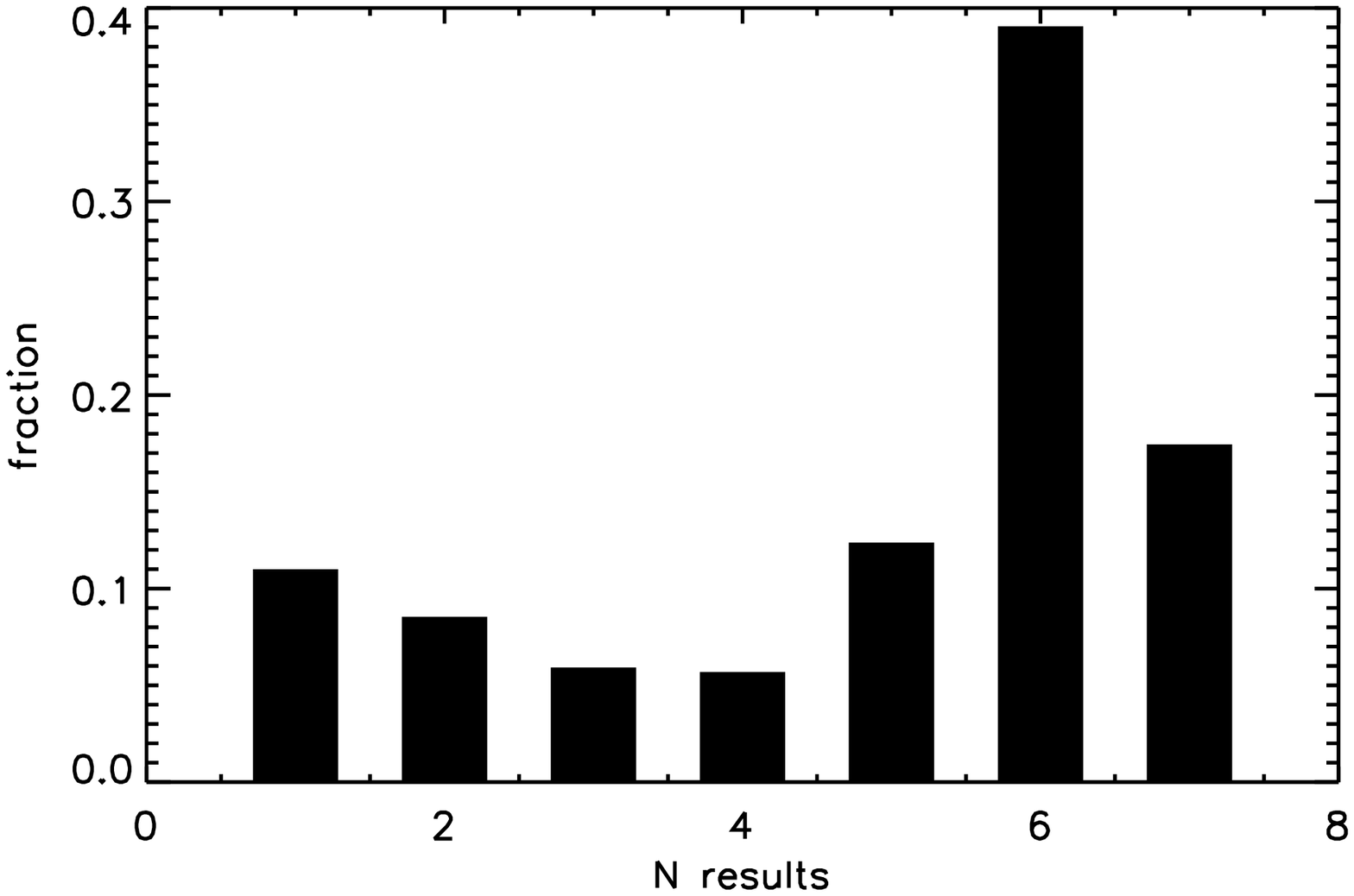}
\end{minipage}
\hfill
\caption{Number of results obtained per star as a function of $\nu_{\rm max}$ (left) and the fractional distribution of the number of results obtained per star (right).}
\label{stat}
\end{figure*}

\begin{figure*}
\begin{minipage}{8.4 cm}
\centering
\includegraphics[width=8.4cm]{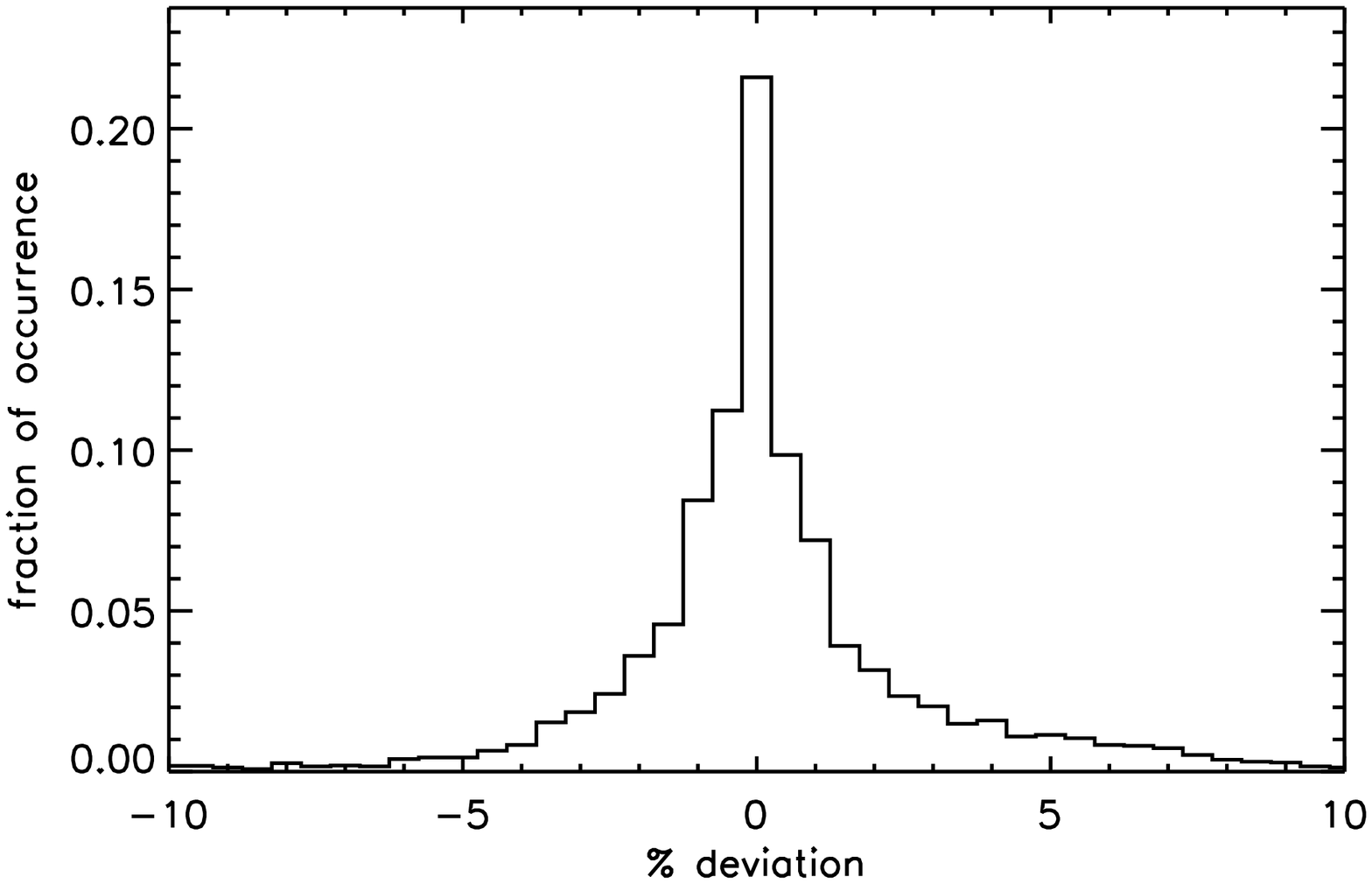}
\end{minipage}
\hfill
\begin{minipage}{8.4cm}
\centering
\includegraphics[width=8.4cm]{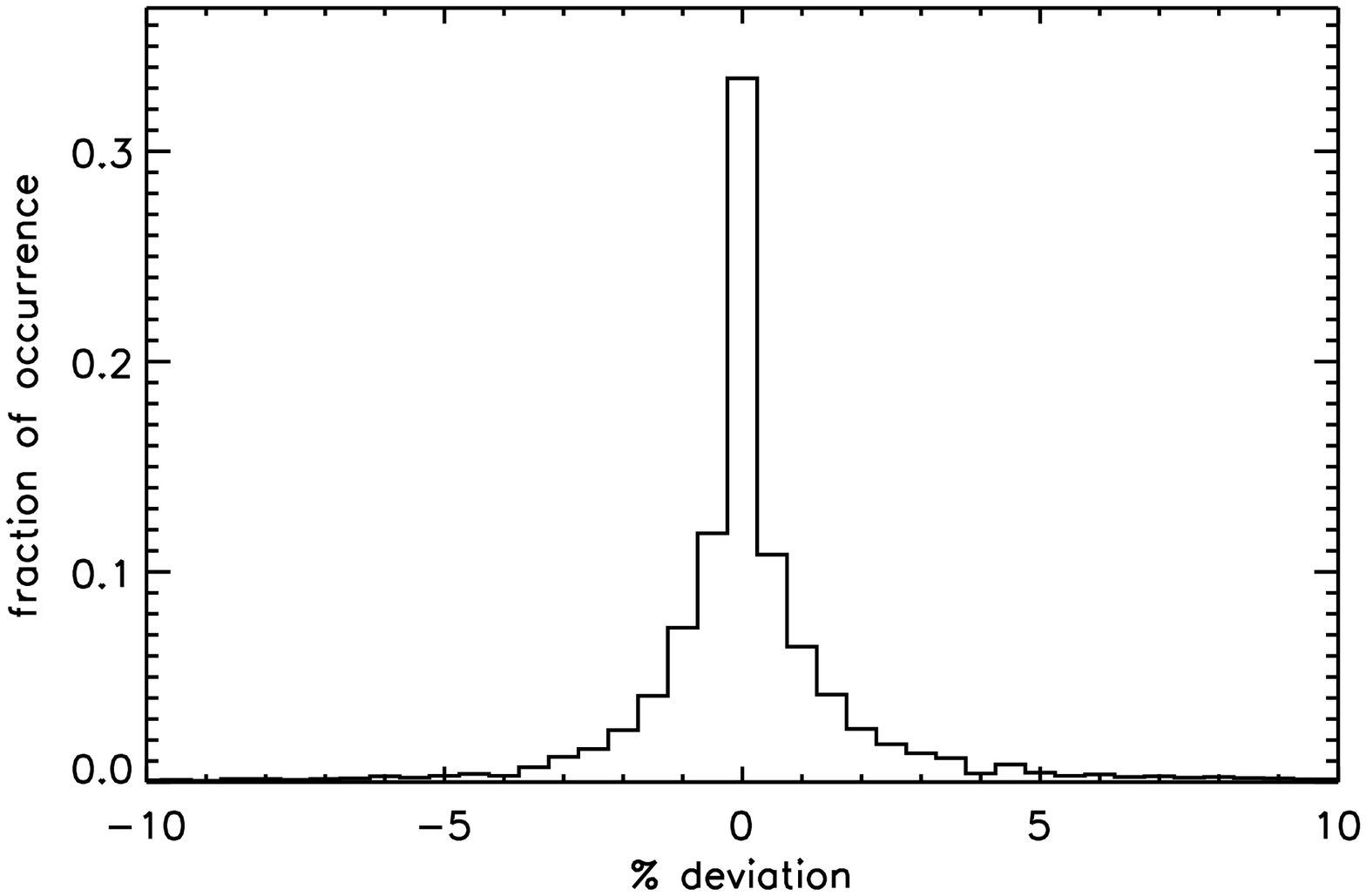}
\end{minipage}
\hfill
\caption{Distributions of the ratio deviation/median value for $\nu_{\rm max}$ (left) and $\Delta \nu$ (right).}
\label{histos}
\end{figure*}

\section{Comparison}
Comparing results from seven different methods that produced seven measures for $\nu_{\rm max}$ and six measures for $\Delta \nu $ for hundreds of stars is a lengthy process. The results of all teams are collected in a data exchange facility called the \textit{Catbasket}. The \textit{Catbasket} is used to store results for individual stars with uncertainties and explanations on how the results were obtained, written in a specific format. The \textit{Catbasket} is located and maintained in Birmingham and is accessible through a web interface. The concept of the \textit{Catbasket}  allows for a comparison of results, as is done in this study.



Most methods have an automated procedure to detect the oscillations. These are based on either the hump of excess power (OCT, DLB) or on the regular frequency pattern of the oscillations (SYD, COR, A2Z). For the CAN method, the stars are selected manually and tend to be those where the signal is strong. The exact implementation of the oscillation detection influences the sensitivity of the methods in any given frequency range. Therefore, the number of methods that return a result varies from star to star. 

Because of the differences between different methods, we resist giving `mean' values for $\nu_{\rm max}$  or $\meandnu$ for any of the \textit{Kepler} stars. However, we do wish to take advantage of the multiple determinations of the parameters to detect and remove values that are discrepant. For this purpose, we compare for each star the results from the individual methods with a median value taken over all results for that star. 
For the detections of outliers we use different definitions for $\nu_{\rm max}$ and $\meandnu$. For the outlier rejection for $\nu_{\rm max}$ an absolute threshold of 15 $\mu$Hz and a relative threshold of 25\% are used. This reflects the roughly constant scatter above 30 $\mu$Hz, but below that level it is more dependent on the median value of $\nu_{\rm max}$ (see top panel of Fig.~\ref{figresults}). This is also justified by the results of the simulations (Sect. 3) where the spread observed in the returned values for $\nu_{\rm max}$ is roughly constant and is less than 5 $\mu$Hz. 
If outlier rejection was applied for $\nu_{\rm max}$ we also excluded the matching value for  $\meandnu$. For the remaining outliers in $\meandnu$ we applied a rejection based on the median absolute deviation (MAD). We use 10$\cdot$MAD unless this is less than 0.5 $\mu$Hz, in which case 0.5 $\mu$Hz is used as the cut-off. This reflects the fact that, for some stars, $\meandnu$ is poorly determined. Note that a minimum of three results is required before outliers can be identified.

\section{Results and discussion}
The results for $\nu_{\rm max}$ and $\meandnu$ from each method are listed in Table~\ref{results} and shown in Fig.~\ref{figresults}. Results have been obtained for 1301 stars. In Fig.~\ref{stat} the number of results obtained per star is shown as a function of $\nu_{\rm max}$ (left), and the fractional distribution of the number of results (right). This histogram shows that for 69\% of the stars we have five or more results, with a clear maximum at six results. For $\nu_{\rm max}$ between roughly 50 and 170 $\mu$Hz there are only very few stars with less than five results, which indicates that the majority of the methods are most sensitive in this frequency interval and less sensitive for oscillations at lower ($<$ 50 $\mu$Hz) and higher ($>$ 170 $\mu$Hz)  frequencies for which fewer methods obtain results. 

We use the median to compare individual results and explore evidence for bias and possible trends of a method compared to the median value of all results. This information can be useful to see how different a method is from the others, but does not by definition tell whether individual methods provide wrong or right values. It is for instance clear that the $\nu_{\rm max}$ values from the first moment of the smoothed power excess (OCT II) are generally higher than results from other methods. Also the spread in these results is relatively large compared to other results. Interestingly, the results of the simulations from the first moment (OCT II) give results that are closer to the input values, i.e. are less spread, than the results from a Gaussian fit to the smoothed power excess (OCT I). It is, however, clear that for the results of the observations compared to the median, the OCT I results are more consistent with the majority of the results, among which the COR, SYD and A2Z results, which also gave results relatively close to the input values of the simulations. Therefore, with the current results of the comparison we can not discard or favour either of the methods OCT I or OCT II, although their results show slightly different behaviour compared to other methods in the simulations and observational results, respectively.  

For $\meandnu$ no significant trends and biases are present in the results from different methods, but the scatter in the results seems to be larger for $\meandnu$ between roughly 3 and 5 $\mu$Hz compared to other values. From population studies \citep[e.g.][]{miglio2009} it is known that stars with $\nu_{\rm max}$ between 30 and 40 $\mu$Hz and $\meandnu$ between 3 and 5 $\mu$Hz are mainly red-clump stars. Due to the fact that stars remain for a relatively long time in the He-burning phase in the red-clump, these stars are most commonly observed \citep{hekker2009,mosser2010}. If we bin the $\meandnu$ values in 1 $\mu$Hz bins we can see that most values are indeed returned between 3 and 5 $\mu$Hz, but that the standard deviation of these values is similar to that in bins at other values of $\meandnu$. So the larger spread in the results in the $\meandnu$ interval 3-5~$\mu$Hz could be an apparent effect caused by the increased number of results. The scatter in the $\Delta \nu$ - $\nu_{\rm max}$ relation (Eq.~\ref{dnunumax}) using results from the different methods is investigated by \citet{huber2010}.

Notwithstanding the differences between the methods and the trends and biases of individual results with respect to the median, the results of the different methods agree within a few percent, as shown in Fig.~\ref{histos}. These figures show the distribution of the ratio of the deviation over the median value of $\nu_{\rm max}$ and $\Delta \nu$ respectively. The deviation and median are correlated values and a bias in a method most likely causes the skewness in the $\nu_{\rm max}$ distribution. 

We also investigate the relation between the absolute deviation from the median and the formal errors, i.e., uncertainties derived within each method, for both $\nu_{\rm max}$ and $\meandnu$. These relations show immediately that for most methods the formal error does not provide a realistic indication of the absolute deviation from the median. Again, this does not tell whether the formal errors are unrealistic, but indicates that for some methods the scatter around the median is considerably different from the computed formal error. Additionally, the formal errors do also not provide an indication of the spread due to realization noise.

\section{Conclusions}
We have performed a comparison of results from different methods for the frequency of maximum oscillation and the mean large frequency separation for over one thousand red-giant stars observed during the first four months of \textit{Kepler} observations. We investigated the differences and consistency between these results and their uncertainties. The effect of realization noise, originating from the stochastic nature of the oscillations, is also investigated using simulated datasets.

The comparison of different realizations of simulated data with the same input values revealed that the scatter due to realization noise is non-negligible and can be at least as important as the internal uncertainty of the result due to the method used, but this depends on the frequency of maximum oscillation power and on the methods, i.e., some methods are more sensitive to realization noise than others. When detailed modelling is performed for red-giant stars, the fact that the observational results are obtained from only one realization has to be taken into account.

For 69\% of the 1301 stars with results, we could obtain $\nu_{\rm max}$ and $\meandnu$ with five or more methods. For stars in a frequency range between 50 and 170 $\mu$Hz this percentage is increased to 92\% indicating that the majority of the methods are sensitive to oscillations in this frequency interval. At lower ($<$ 50 $\mu$Hz) frequencies less results are obtained. This is firstly due to the limited frequency resolution of the data and secondly due to difficulties in disentangling oscillations from other effects, such as granulation or instrumental effects, which are also present at low frequencies. At higher ($>$ 170 $\mu$Hz) frequencies the lower height, i.e. lower signal-to-noise ratio, of the oscillations causes more problems for some methods to detect the oscillations. Furthermore, the Nyquist frequency of these data is $\sim$~280~$\mu$Hz, which possibly causes difficulties  due to reflection effects, while for some stars only part of the oscillation power excess is covered. In these cases the determination of the background signal is also more complicated. 

Although some biases and trends with respect to the median value of the results are present in each method, the results from the different methods agree for most stars within a few percent. Despite the consistency between the results from different methods, the stated formal errors are not indicative of the deviation of these results from the median value.

For $\meandnu$, we found that the results do not show significant dependence on the range over which they were calculated. This is plausible because the trend in $\Delta\nu$ is approximately linear with frequency and/or shows rapid variations with relatively low amplitude \citep{mosser2010}, which average out. This is the case for the red giants investigated here with oscillations over a range of roughly 10 to 280 $\mu$Hz.

Apart from the results described here, the comparison also helped to improve all the methods and will continue to do so. Although we are dealing with automated methods, it appeared that inspection of the results is important. In order to remove false positives, it remains necessary to inspect the results in specific frequency ranges by eye: at low frequencies, close to the Nyquist frequency or in regions of known artefacts as described in Sect. 4. An additional inspection has been performed for stars for which the results of different methods were significantly different. The latter inspection revealed in most cases that different features were selected as being the oscillations. In some cases this could be traced back to artefacts, as discussed in Sect. 4, or contamination of background stars. 

To summarise, the comparison between results of global oscillation parameters from different methods has been very useful, because it allowed for the first time to provide a qualitative measure of the differences between different methods. The comparison has been used to investigate the contribution of the realization noise and internal uncertainties in the methods and allowed for the detection of artefacts. These issues will remain important and the comparison will continue to be applied. In the future, we will investigate the improvements in the accuracy of the results when we have data with longer timespan and investigate the actual uncertainties in the results.

\acknowledgements
Funding for the \textit{Kepler Mission} is provided by NASA's Science Mission Directorate. The authors gratefully acknowledge the \textit{Kepler} Science Team and all those who have contributed to making the \textit{Kepler Mission} possible. SH, YE, SJH and WJC acknowledge financial support from the UK Science and Technology Facilities Council (STFC).
The research leading to these results has received funding from the European
Research Council under the European Community's Seventh Framework Programme
(FP7/2007--2013)/ERC grant agreement n$^\circ$227224 (PROSPERITY), as well as from
the Research Council of K.U.Leuven grant agreement GOA/2008/04. The National Center for  
Atmospheric Research is a federally funded research and development  
center sponsored by the U.S. National Science Foundation.
DH acknowledges support by the Astronomical Society of Australia (ASA).

\bibliographystyle{aa}
\bibliography{15185bib}

\longtab{7}{
\begin{longtable}{c|cc|cc|cc|cc|cc|ccc}
\caption{\label{results} Results for $\nu_{\rm max}$ and $\Delta \nu$ from different methods. Up-to-date results can be requested from the authors.}\\
\hline\hline
 & COR & COR & CAN & CAN & A2Z & A2Z & SYD & SYD & DLB & DLB & OCT I & OCT II & OCT\\
star & $\nu_{\rm max}$ & $\Delta \nu$ & $\nu_{\rm max}$ & $\Delta \nu$ & $\nu_{\rm max}$ & $\Delta \nu$ & $\nu_{\rm max}$ & $\Delta \nu$ & $\nu_{\rm max}$ & $\Delta \nu$ & $\nu_{\rm max}$ & $\nu_{\rm max}$ & $\Delta \nu$\\
 & $\mu$Hz & $\mu$Hz & $\mu$Hz & $\mu$Hz & $\mu$Hz & $\mu$Hz & $\mu$Hz & $\mu$Hz & $\mu$Hz & $\mu$Hz & $\mu$Hz & $\mu$Hz &$\mu$Hz\\
\hline
\endfirsthead
\caption{Continued.}\\
\hline\hline
 & COR & COR & CAN & CAN & A2Z & A2Z & SYD & SYD & DLB & DLB & OCT I & OCT II & OCT\\
star & $\nu_{\rm max}$ & $\Delta \nu$ & $\nu_{\rm max}$ & $\Delta \nu$ & $\nu_{\rm max}$ & $\Delta \nu$ & $\nu_{\rm max}$ & $\Delta \nu$ & $\nu_{\rm max}$ & $\Delta \nu$ & $\nu_{\rm max}$ & $\nu_{\rm max}$ & $\Delta \nu$\\
 & $\mu$Hz & $\mu$Hz & $\mu$Hz & $\mu$Hz & $\mu$Hz & $\mu$Hz & $\mu$Hz & $\mu$Hz & $\mu$Hz & $\mu$Hz & $\mu$Hz & $\mu$Hz &$\mu$Hz\\
\hline
\endhead
\hline
\endfoot
1161618  &        33.50 &         4.22 &        33.58 &         4.15 &        33.79 &   &        34.02 &         4.12 &        32.84 &         3.55 &        34.05 &        34.86 &         4.18   \\ 
1162746  &        27.20 &         3.80 &        26.39 &         4.10 &   &   &        27.68 &   &   &   &        27.90 &        28.80 &     \\ 
1432587  &         4.30 &         0.91 &   &   &   &   &   &   &   &   &   &   &     \\ 
1433593  &        69.00 &         6.31 &        68.98 &         6.17 &        69.19 &         6.35 &        68.81 &         6.22 &   &   &        70.06 &        71.23 &         6.34   \\ 
1433730  &        41.40 &         4.12 &        41.58 &         4.16 &        38.17 &         4.03 &        41.45 &         4.16 &        37.57 &         4.38 &        42.02 &        44.73 &         4.11   \\ 
1435573  &        26.80 &         3.76 &        26.72 &         3.83 &   &   &        27.27 &         3.64 &   &   &        27.12 &        28.00 &         3.67   \\ 
1719297  &         7.60 &         1.19 &   &   &   &   &         7.59 &   &   &   &   &   &     \\ 
1720425  &        28.00 &         3.60 &   &   &        28.26 &         3.60 &        28.31 &         3.70 &        26.87 &         3.49 &        28.65 &        29.84 &         3.60   \\
\end{longtable}
}

\end{document}